\documentclass[english,12pt,aps,prd,a4paper,preprintnumbers,floatfix,nofootinbib,showpacs,superscriptaddress, notitlepage]{revtex4-1} 
 \pdfoutput=1
\usepackage[usenames,dvipsnames]{color}  
\usepackage{graphicx}
\usepackage{caption}
\usepackage{subcaption}
\usepackage[export]{adjustbox} 
\usepackage{amsmath}
\usepackage{amssymb}
\usepackage[colorlinks=true,citecolor=darkred,urlcolor=darkred, pdfborder={0 0 0}]{hyperref}
\usepackage[normalem]{ulem}
\usepackage{braket}

\makeatletter
\def\p@subsection{}
\makeatother

\definecolor{darkred}{rgb}{0.6,0,0}

\definecolor{linkcolor}{rgb}{0,0,0.5}

\def\gsim{\raise0.3ex\hbox{$\;>$\kern-0.75em\raise-1.1ex\hbox{$\sim\;$}}}
\def\lsim{\raise0.3ex\hbox{$\;<$\kern-0.75em\raise-1.1ex\hbox{$\sim\;$}}}

\def\beqn#1{\begin{equation}\label{#1}}
\def\eeqn{\end{equation}}

\def\beqa#1{\begin{eqnarray}\label{#1}}
\def\eeqa{\end{eqnarray}}

\def\znbb {neutrinoless double beta }
\def\0nbb {$0\nu\beta\beta$ }

\def\Z2{$\mathcal{Z_2}$}

\def\black{\color{black}{}}

\newcommand {\ignore}[1]{}

\newcommand{\sm}{{Standard Model }}
\def\lfv{lepton flavour violation }

\def\lfv{lepton flavour violation }

\def\SM{$\mathrm{SU(3)_c \otimes SU(2)_L \otimes U(1)_Y}$ }

\def\321{$\mathrm{SU(3) \otimes SU(2) \otimes U(1)}$ }

\newcommand{\AddrAHEP}{%
  AHEP Group, Institut de F\'{i}sica Corpuscular --
  CSIC/Universitat de Val\`{e}ncia, Parc Cient\'ific de Paterna.\\
 C/ Catedr\'atico Jos\'e Beltr\'an, 2 E-46980 Paterna (Valencia) - SPAIN}

\newcommand{\AddrIISERB}{Department of Physics, Indian Institute of Science Education and Research - Bhopal, \\ 
Bhopal Bypass Road, Bhauri, Bhopal 462066, India}

\begin{document}

\bibliographystyle{unsrt}   

\title{\color{BrickRed}  
The simplest scoto-seesaw model: \\WIMP dark matter phenomenology and Higgs vacuum stability}

\author{Sanjoy Mandal}\email{smandal@ific.uv.es}
\affiliation{\AddrAHEP}
\author{Rahul Srivastava}\email{rahul@iiserb.ac.in}
\affiliation{\AddrIISERB}
\author{Jos\'{e} W. F. Valle}\email{valle@ific.uv.es}
\affiliation{\AddrAHEP}

\begin{abstract}
  \vspace{1cm} 

 We analyze the consistency of electroweak breaking, neutrino and dark matter phenomenology within the simplest scoto-seesaw model.
By adding the minimal dark sector to the simplest ``missing partner'' type-I seesaw one has a physical picture for the neutrino oscillation lengths:
the ``atmospheric'' mass scale arises from the tree-level seesaw, while the ``solar'' scale is induced radiatively, mediated by the dark sector.
We identify parameter regions consistent with theoretical constraints, as well as dark matter relic abundance and direct detection searches.
Using two-loop renormalization group equations we explore the stability of the vacuum and the consistency of the underlying dark parity symmetry. 
One also has a lower bound for the \znbb decay amplitude.

\end{abstract}

\maketitle


\section{Introduction}
\label{sec:introduction}

The discovery of neutrino oscillations~\cite{McDonald:2016ixn,Kajita:2016cak} implies that at least two neutrinos are massive.
There has by now been strong evidence, at different scales, for the existence of cosmological dark matter, the basic understanding and interpretation of which we also lack~\cite{Bertone:2004pz}.
The main current neutrino mass generation paradigms are the seesaw and the scotogenic mechanism,
which also accounts for dark matter as the mediator of neutrino mass, as a result of an assumed $\mathbb{Z}_2$ symmetry.
Both mechanisms give mass ``democratically'' to all neutrino states, according to the structure of the relevant Yukawa couplings.

The simplest ``scoto-seesaw'' extension of the Standard Model~\cite{Rojas:2018wym} combines these two main paradigms within its minimal \SM framework.
In such hybrid scenario the atmospheric scale comes from the tree level seesaw, while the solar scale is mediated by the radiative exchange of dark states, i.e.
\begin{align}
 \Delta m_{\text{ATM}}^{2}= \left(\frac{v^{2}}{2M_{N}}\mathbb{Y}_{N}^{2}\right)^{2},\,\,\, 
 \Delta m_{\text{SOL}}^{2}\approx \left(\frac{1}{32\pi^{2}}\right)^{2}\left(\frac{\lambda_{5}v^{2}}{M_{f}^{2}-m_{\eta^{}}^{2}}M_{f}\mathbb{Y}_{f}^{2}\right)^{2},
 \label{eq:sol-atm}
\end{align}
where $M_{N}$ is the ``right-handed'' neutrino mass, $M_{f}$  and $m_{\eta^{}}$ are ``dark-sector'' masses
and $\mathbb{Y}_{N}^2$,~$\mathbb{Y}_{f}^2$ are corresponding Yukawa coupling strengths.

One sees that the solar splitting will be non-zero as long as $\lambda_{5} \neq 0$.
Moreover, one accounts naturally for the  hierarchy between the solar and atmospheric scales observed in the experimental data~\cite{deSalas:2020pgw}.  The
corresponding ratio of squared solar-to-atmospheric mass splittings for normal and inverted mass hierarchy are found to be~\cite{10.5281/zenodo.4593330,deSalas:2020pgw}:
\begin{align}
\textbf{NO:}\, \frac{\Delta m_{\text{SOL}}^{2}}{\Delta m_{\text{ATM}}^{2}}=0.0294^{+0.0027}_{-0.0023},\,\,\,\,\textbf{IO:}\,\frac{\Delta m_{\text{SOL}}^{2}}{\Delta m_{\text{ATM}}^{2}}=0.0306^{+0.0028}_{-0.0025}.
  \label{eq:sol-atm-obs}
\end{align} 
Altogether, the interplay of the ``seesaw'' and ``dark-sectors'' provide an interesting way to describe lepton number violation and neutrino mass generation.
Indeed, the scoto-seesaw model has a viable weakly interacting massive particle (WIMP) dark matter candidate and accounts for the observed neutrino masses, including the solar-to-atmospheric hierarchy.
The aim of this work is to explore the scoto-seesaw model in more detail.

The paper is organized as follows. In section~\ref{sec:scoto} we briefly describe the model, giving the details of the new fields and their interactions. 
In section~\ref{sec:Neutrino Masses}, we describe the tree level and radiative neutrino mass generation.
In section~\ref{sec:dark matter} we study the parameter space for the case of a scalar dark matter candidate.
In section~\ref{sec:vacuum stability} we look at vacuum stability in the scoto-seesaw model and show that, over large parameter regions, the vacuum is stable all the way up to the Planck scale.
In section~\ref{sec:dark-parity} we examine the robustness of the dark parity symmetry under renormalization group (RG) evolution of the parameters. 
We finally conclude in section~\ref{sec:summary-discussion}.

\section{Minimal scoto-seesaw model}
\label{sec:scoto}

The minimal combination of the seesaw mechanism and the scotogenic model was proposed in Ref.~\cite{Rojas:2018wym}
\footnote{One can have scoto-seesaw realizations based on (3,2) seesaw extensions~\cite{Barreiros:2020gxu}.
  While they have new interesting features, one looses the interesting prediction in Eq.~\eqref{eq:sol-atm}.}.
It clones the simplest ``missing partner'' (3,1) version of the Standard Model seesaw mechanism suggested in~\cite{Schechter:1980gr,Schechter:1981cv}
with the minimal scotogenic model proposed in~\cite{Ma:2006km}.
We now describe in detail both the fermionic and the scalar sectors of the model. 

\subsection{The Yukawa Sector}
\label{Yukawa Sector}

We now briefly recall the basic features of the minimal scoto-seesaw model~\cite{Rojas:2018wym}. The new particles and their charges are given in Table~\ref{tab:quantun-numbers},
where the family index $a$ runs from 1 to 3.
%
%
\begin{table}[!t]
\centering
\begin{tabular}{|c||c|c|c||c|c||c|c|}
\hline
        & \multicolumn{3}{|c||}{Standard Model} &  \multicolumn{2}{|c||}{New Fermions}  & \multicolumn{1}{|c|}{New Scalar}  \\
        \cline{2-7}
        & \hspace{0.25cm} $L_a$  \hspace{0.25cm} &  \hspace{0.25cm} $e_a$  \hspace{0.25cm} &  \hspace{0.25cm}$H$ \hspace{0.25cm}  &  \hspace{0.2cm} $N$ \hspace{0.2cm} &  \hspace{0.1cm} $f$  \hspace{0.1cm}  &  \hspace{0.01cm}$\eta$   \hspace{0.01cm}\\
\hline     
$SU(2)_L$ &  2    &  1    &    2    &     1    &  1   &    2       \\
\hline
$U(1)_Y$     & -1/2    &  -1    &    1/2    &     0    &  0   &    1/2    \\
\hline
$\mathbb{Z}_2$   &  $+$  &  $+$  &   $+$  &  $+$   & $-$  &  $-$     \\
\hline
\end{tabular}
\caption{Matter content and charge assignment of the minimal scoto-seesaw model.}
\label{tab:quantun-numbers}
\end{table}

In Table.~\ref{tab:quantun-numbers} the additional $\mathbb{Z}_2$ symmetry is the ``dark parity'' responsible for the stablity of the dark matter candidate. 
All the \sm particles and $N$ are even under this dark $\mathbb{Z}_{2}$ parity,
while the dark sector, consisting of one fermion $f$ and one scalar $\eta$, is odd under $\mathbb{Z}_{2}$.

The full Yukawa sector can be split as
\begin{align}
 \mathcal{L}=\mathcal{L}_{\text{SM}}+\mathcal{L}_{\text{ATM}}+\mathcal{L}_{\text{DM,SOL}}
\end{align}
where $\mathcal{L}_{\text{SM}}$ is the Standard Model Lagrangian, while
\begin{align}
 \mathcal{L}_{\text{ATM}}= -  Y_{N}^{a}\bar{L}^{a}  \tilde{H} N  + 
 \frac{1}{2}  M_{N} \overline{N^{c}}N + h.c,
\label{ATM Lagrangian}
\end{align}
induces the type-I seesaw neutrino mass (atmospheric neutrino mass scale) after the electroweak symmetry breaking. Note also that throughtout this work repeated indices imply summation,
with $\tilde{H} = i\sigma_{2}H^{*}$, $\sigma_2$ being the second Pauli matrix.

The Lagrangian responsible for the solar and dark sector is given by
\begin{align}
 \mathcal{L}_{\text{DM,SOL}} = Y_{f}^{a}\bar{L}^{a} \tilde{\eta}~f+\frac{1}{2}M_{f}\overline{f^{c}}f+h.c.
\label{SOL Lagrangian}
\end{align}
It induces the solar neutrino mass scale as discussed in Sec.~\ref{sec:Neutrino Masses}.

\subsection{The Scalar Sector}
\label{Scalar Sector}

Apart from the Standard Model (SM) Higgs doublet $H$ we have a scalar doublet $\eta$ carrying the same quantum numbers, but with $\mathbb{Z}_2$-odd parity.
The \SM gauge invariant scalar potential is given by
\begin{align}
 V&=-\mu_{H}^{2}H^{\dagger}H+m_{\eta}^{2}\eta^{\dagger}\eta+\lambda (H^{\dagger}H)^{2}+\lambda_{\eta}(\eta^{\dagger}\eta)^{2}
 +\lambda_{3}(H^{\dagger}H)(\eta^{\dagger}\eta)+\lambda_{4}(H^{\dagger}\eta)(\eta^{\dagger}H)\nonumber \\
 &+\frac{\lambda_{5}}{2}\left((H^{\dagger}\eta)^{2}+h.c.\right)
 \label{eq:scoto-pot}
\end{align}

We now turn to the consistency conditions of the potential. 
The following restrictions must hold so as to ensure that the scalar potential is bounded from below and has a stable vacuum at any given energy scale $\mu$: 
\begin{align}
 \lambda (\mu) >0,\,\,\lambda_{\eta}(\mu)>0,
\label{stability1}
\end{align}
\begin{align}
 \lambda_A\equiv\lambda_3 (\mu)+\sqrt{4\lambda(\mu)\lambda_{\eta}(\mu)} > 0,
\label{stability2}
\end{align}
\begin{align}
\lambda_B\equiv\lambda_3(\mu)+\lambda_4(\mu)+\sqrt{4\lambda(\mu)\lambda_\eta(\mu)}-|\lambda_5(\mu)| > 0.
\label{stability3}
\end{align}
%
where $\lambda_i(\mu)$ are the values of the quartic couplings at the running scale $\mu$.
In order to have an absolutely stable vacuum, one must satisfy the conditions given in Eqs.~(\ref{stability1}), (\ref{stability2}) and (\ref{stability3}) at each and every energy scale.
To ensure perturbativity, we take a conservative approach of simply requiring that the scalar quartic couplings in Eq.~\eqref{eq:scoto-pot} obey $\leq 4\pi$.

\subsection{Mass Spectrum} 

In order to ensure dark matter stability the $\mathbb{Z}_2$ symmetry should remain unbroken.
  This means that the $\mathbb{Z}_2$ odd scalar $\eta$ should not acquire a nonzero vacuum expectation value (VEV).
As a result, electroweak symmetry breaking is driven simply by the VEV of $H$. The fields $\eta$ and $H$ can be expanded as follows 
\begin{align}
H=
\begin{pmatrix}
H^+\\
(v+h+i\phi^0)/\sqrt{2}
\end{pmatrix},\,\,
\eta=
\begin{pmatrix}
\eta^+\\
(\eta^R+i\eta^I)/\sqrt{2}
\end{pmatrix}
\end{align}
Exact conservation of the $\mathbb{Z}_{2}$ symmetry forbids the mixing between the Higgs and the dark doublet $\eta$.
The components of $\eta$ have the following masses
\begin{align}
 m_{\eta^{R}}^{2}&=m_{\eta}^{2}+\frac{1}{2}\left(\lambda_{3}+\lambda_{4}+\lambda_{5}\right)v^{2}\\
  m_{\eta^{I}}^{2}&=m_{\eta}^{2}+\frac{1}{2}\left(\lambda_{3}+\lambda_{4}-\lambda_{5}\right)v^{2}\\
 m_{\eta^{+}}^{2}&=m_{\eta}^{2}+\frac{1}{2}\lambda_{3}v^{2}.
\end{align}
The difference $m_{\eta^R}^2-m_{\eta^I}^2$ depends only on the parameter $\lambda_5$ which, we will show later, is also responsible for smallness of solar neutrino mass scale.  
The conservation of the $\mathbb{Z}_2$ symmetry also makes the lightest of the two eigenstates $\eta^R$ and $\eta^I$ a viable scalar dark matter candidate\footnote{Throughout this work we assume that the dark fermion $f$ is heavier than the dark scalars $\eta^R$ and $\eta^I$.}, as we explore in what follows.
\section{Neutrino Masses} 
\label{sec:Neutrino Masses}

At tree-level this model gives rise to the following neutrino mass matrix $\mathcal{M}_\nu^{ab}$,
\begin{align}
\mathcal{M}_\nu^{ab}=
\begin{pmatrix}
0 & 0 & 0 & \frac{Y_N^1 v}{\sqrt{2}} \\
0 & 0 & 0 & \frac{Y_N^2 v}{\sqrt{2}} \\
0 & 0 & 0 & \frac{Y_N^3 v}{\sqrt{2}} \\
\frac{Y_N^1 v}{\sqrt{2}} & \frac{Y_N^2 v}{\sqrt{2}} & \frac{Y_N^3 v}{\sqrt{2}} & M_N \\
\end{pmatrix}
\end{align}
in the basis $(L^a, N)^T$. Notice the (3,1) structure of the seesaw~\cite{Schechter:1980gr,Schechter:1981cv},
as a result of which one sees that the $N$ pairs off with one combination of the doublets in $L_a$ through their Dirac-like couplings.

This clearly leads to a projective structure for the tree-level neutrino mass matrix
\begin{align}
\mathcal{M}_{\nu\text{TREE}}=-\frac{v^2}{2M_N}Y_N^a Y_N^b,
\label{atmospheric scale}
\end{align} 
where $a,b=1,2,3$ are family indices of the lepton doublets.
One sees from Eq.~\eqref{atmospheric scale} that for ``sizeable'' Yukawa couplings, $Y_N\sim\mathcal{O}(1)$, in order to reproduce the required value of the atmospheric scale,
heavy neutrinos must lie at mass scale $M_N\sim\mathcal{O}(10^{14}\,\text{GeV})$.
Smaller values of the Yukawa coupling $Y_N$ would require correspondingly lower seesaw scale $M_N$.  

\begin{figure}[h]
\centering
\includegraphics[width=0.4\textwidth]{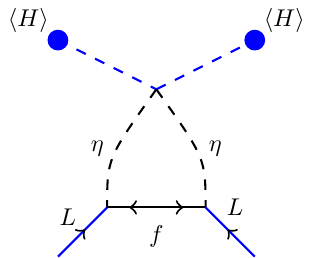}
\caption{Solar neutrino mass scale induced from radiative dark-sector exchange.} 
\label{fig:solar-loop}
\end{figure}

The solar mass scale arises from Fig.~\ref{fig:solar-loop} involving the exchange of the scalar and fermionic dark mediators $\eta$ and $f$.
Although these corrections also have a projective structure, they break the ``missing-partner'' nature of the (3,1) type-I seesaw mechanism\footnote{The situation is analogous to neutrino mass generation in bilinear broken R-parity supersymmetry~\cite{Hirsch:2000ef,Diaz:2003as,Hirsch:2004he}.}.
The total neutrino mass has the form
\begin{equation}
 \mathcal{M}_{\nu \text{TOT}}^{ab}=-\frac{v^{2}}{2M_{N}}Y_{N}^{a}Y_{N}^{b}+\mathcal{F}(m_{\eta^{R}},m_{\eta^{I}},M_{f}) M_{f}Y_{f}^{a}Y_{f}^{b},
 \label{eq:mnu-scoto-seesaw}
\end{equation}
where the first term is the tree-level seesaw part, and the loop function $\mathcal{F}$ characterizes the quantum correction arising from Fig.~\ref{fig:solar-loop}.
This is responsible for inducing the solar mass scale.

The loop function $\mathcal{F}$ is expressed as the difference of two $B_{0}$-Veltman functions, namely,
\begin{align}
 \mathcal{F}(m_{\eta^{R}},m_{\eta^{I}},M_{f})=\frac{1}{32\pi^{2}} \left(\frac{m_{\eta^{R}}^{2}\,\text{log}\left(M_{f}^{2}/m_{\eta^{R}}^{2}\right)}{M_{f}^{2}-m_{\eta^{R}}^{2}}-
 \frac{m_{\eta^{I}}^{2}\,\text{log}\left(M_{f}^{2}/m_{\eta^{I}}^{2}\right)}{M_{f}^{2}-m_{\eta^{I}}^{2}}\right)
\end{align} 
Since both terms in Eq.~\eqref{eq:mnu-scoto-seesaw} have a projective nature, one out of the three neutrinos remains massless.

From the eigenvalues of the neutrino mass matrix $\mathcal{M}_{\nu}^{ab}$ one can estimate the atmospheric and solar square mass differences as,
\begin{align}
 \Delta m_{\text{ATM}}^{2}= \left(\frac{v^{2}}{2M_{N}}\mathbb{Y}_{N}^{2}\right)^{2},\,\,\, 
 \Delta m_{\text{SOL}}^{2}\approx \left(\frac{1}{32\pi^{2}}\right)^{2}\left(\frac{\lambda_{5}v^{2}}{M_{f}^{2}-m_{\eta^{R}}^{2}}M_{f}\mathbb{Y}_{f}^{2}\right)^{2}.
 \label{atm-sol-mass}
\end{align}
where we take $M_{f}^{2}$, $m_{\eta^{R}}^{2}$, $M_{f}^{2}-m_{\eta^{R}}^{2}\gg\lambda_{5}v^{2}$ and $\mathbb{Y}_{\ell}^{2}=(Y_{\ell}^{e})^{2}+(Y_{\ell}^{\mu})^{2}+(Y_{\ell}^{\tau})^{2}$ for $\ell=N,\,f$.
It follows that the ratio between the solar and atmospheric square mass differences can be written as:
\begin{align}
\frac{\Delta m_{\text{SOL}}^{2}}{\Delta m_{\text{ATM}}^{2}}\approx \left(\frac{1}{16\pi^2}\right)^2  \left(\lambda_5\frac{M_N M_f}{M_f^2-m_{\eta^R}^2}\right)^2 \left(\frac{\mathbb{Y}_f^2}{\mathbb{Y}_N^2}\right)^2
\label{eq:ratio}
\end{align}
From Eq.~\eqref{atm-sol-mass} it is clear that one can fit the observed atmospheric and solar mass square differences 
in many ways as long as one takes an adequately small value for $\lambda_5$.
Moreover, Eq.~\eqref{eq:ratio} nicely reproduces Eq.~\eqref{eq:sol-atm-obs}.
In the following we list some choices which can satisfy both the solar and atmospheric scales, as well as have $\eta^R$ as the scalar WIMP dark matter: 
\begin{itemize}
\item  $M_N\sim 10^{14}$ GeV, $M_f\sim 10^{12}$ GeV, $m_{\eta^R}\sim 10^3$ GeV, $\mathbb{Y}_N\sim 0.4$, $\mathbb{Y}_f\sim 0.4$,
\item  $M_N\sim 10^{12}$ GeV, $M_f\sim 10^{4}$ GeV, $m_{\eta^R}\sim 10^3$ GeV, $\mathbb{Y}_N\sim 0.1$, $\mathbb{Y}_f\sim 10^{-4}$,
\item  $M_N\sim 10^{14}$ GeV, $M_f\sim 10^{5}$ GeV, $m_{\eta^R}\sim 10^3$ GeV, $\mathbb{Y}_N\sim 0.4$, $\mathbb{Y}_f\sim 10^{-4}$,
\item  $M_N\sim 10^{6}$ GeV, $M_f\sim 10^{6}$ GeV, $m_{\eta^R}\sim 10^3$ GeV, $\mathbb{Y}_N\sim 10^{-5}$, $\mathbb{Y}_f\sim 10^{-4}$.
\end{itemize}
The upshot of this discussion is that one can easily fit the solar and atmospheric scales for reasonable parameter choices. 
For example, for sufficiently small $\lambda_5$ values, one can choose a reasonable Yukawa coupling $\mathbb{Y}_f$ and $M_f\gsim\mathcal{O}$(TeV).
In section~\ref{sec:dark matter} we show that either $\eta^I$ or $\eta^R$ can, indeed, be taken as a consistent WIMP dark matter candidate.

\section{Phenomenology of Scalar WIMP Dark Matter} 
\label{sec:dark matter}

In this section we collect the results of our analysis of dark matter phenomenology.
In addition to ensuring radiative generation of neutrino masses, the $\mathbb{Z}_2$ symmetry in the dark sector ensures the stability of ``lightest dark particle'' (LDP).
Such LDP is in principle a viable dark matter candidate. There are three LDP options.
The first is the dark fermion $f$. 
The others are the real and imaginary parts of the $\eta$ scalar, $\eta^R$ and $\eta^I$.
In our analysis, we assume scalar dark matter, with the condition $\lambda_5<0$ on the quartic coupling $\lambda_5$. As a result $\eta^R$ will be our dark matter candidate 
(the opposite scenario with $\lambda_5>0$ would have $\eta^I$ as the dark matter particle).

In order to calculate all the vertices, mass matrices, tadpole equations etc the model is implemented in the SARAH package~\cite{Staub:2015kfa}. 
On the other hand, the thermal component of the dark matter relic abundance, as well as the dark matter-nucleon scattering cross section, are determined using micrOMEGAS-5.0.8~\cite{Belanger:2018ccd}.

\subsection{Relic density}  

As shown in Fig.~\ref{fig:annihiliation} (Appendix~\ref{AppendixB}), there are several dark matter annihilation and coannihilation diagrams present in the scoto-seesaw model.
They involve annilation to quarks and leptons, SM gauge bosons and the Higgs boson.
Altogether, they determine the relic abundance of our assumed LDP, $\eta^R$. 
Our numerical scan is performed varying the input parameters as given in Table~\ref{tab:scan}. 
\begin{table*}[ht]
	\centering
\begin{tabular}{|c|c|}
\hline
Parameters & Range \\
\hline
$m_\eta^2$ & $[100^2,5000^2]\,(\text{GeV}^2)$ \\
$\lambda_3$ & $[10^{-5},1]$ \\
$\lambda_4$ & $[10^{-5},1]$ \\
$|\lambda_5|$ & $[10^{-5},10^{-3}]$ \\
\hline
\end{tabular}
\caption{Ranges of variation of the input parameters used in our numerical scan.}
 \label{tab:scan}
\end{table*} 

\begin{figure}[h]
\centering
\includegraphics[width=0.8\textwidth]{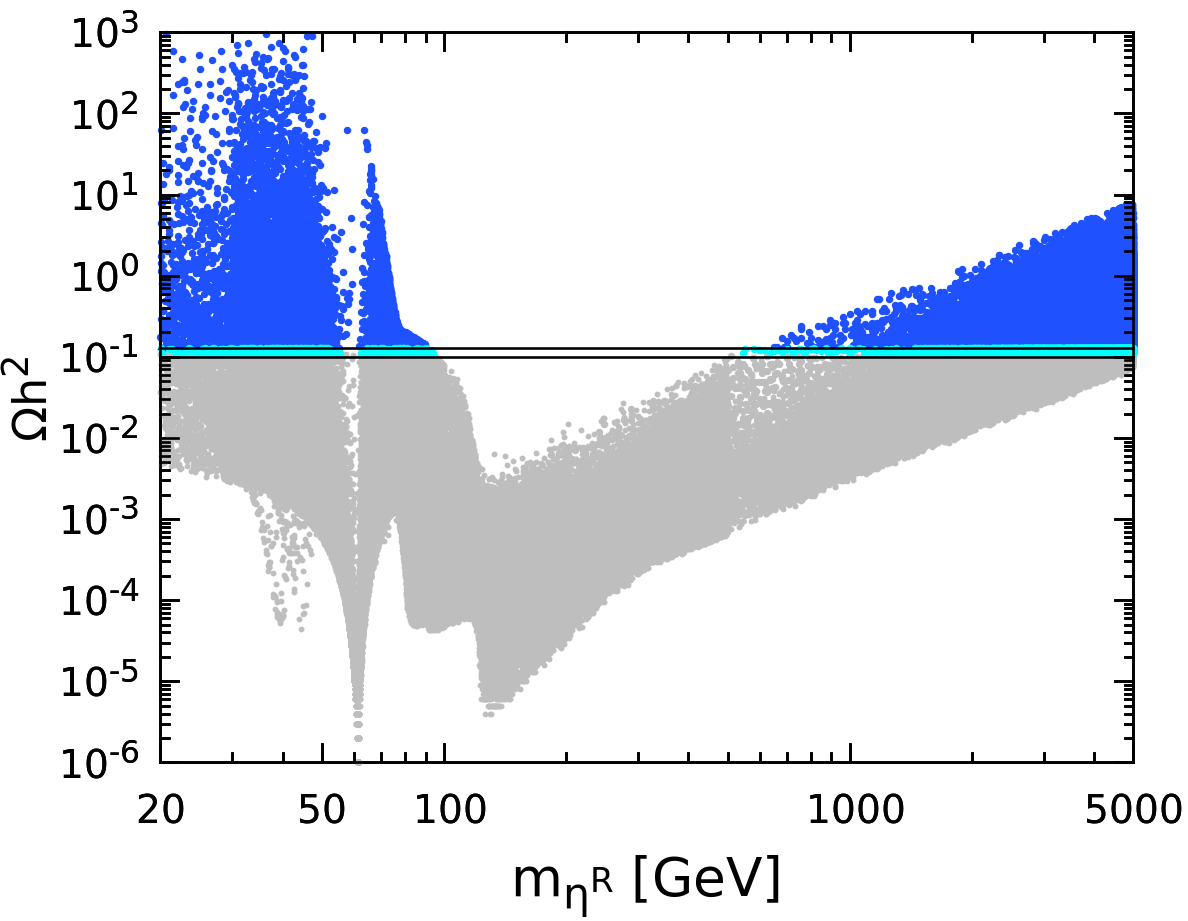}
\caption{\footnotesize{ Relic abundance as a function of the dark matter mass $m_{\eta^R}$.
    Cyan points inside the black lines fall within the measured $3\sigma$ cold dark matter relic density range given by Planck satellite data, Eq.~\eqref{eq:Pl}.
    Gray (blue) points outside the narrow band give under (over) abundance of dark matter, respectively.}}
\label{fig:Relic}
\end{figure}

In Fig.~\ref{fig:Relic} we show the relic density as a function of the mass of the scalar dark matter candidate $\eta^R$.
The narrow horizontal band is the $3\sigma$ range for cold dark matter derived from the Planck satellite data~\cite{Aghanim:2018eyx}:
\begin{align}
  \label{eq:Pl}
0.1126 \leq \Omega_{\eta^R} h^2 \leq 0.1246.
\end{align}
Only for solutions falling exactly within this band the totality of the dark matter can be explained by $\eta^R$.  
The relic density for the cyan points in Fig.~\ref{fig:Relic} lies within the above $3\sigma$ range, whereas the relic density for blue and gray points is above and below the $3\sigma$ range. 
One sees from Fig.~\ref{fig:Relic} that the correct relic density can be obtained in three mass ranges: $m_{\eta^R}<50$~GeV, $70\,\text{GeV}< m_{\eta^R}< 100$~GeV and $m_{\eta^R}>550$~GeV.
The reasons for these mass gaps can be understood by looking in detail into the $\eta^R$ annihilation channels (see Appendix~\ref{AppendixB}). 
The first dip occurs at $m_{\eta^R}\sim M_Z/2$ and corresponds to annihilation via s-channel $Z$ exchange.
The second depletion of the relic density happens around $m_{\eta^R}\sim m_h/2$ and corresponds to annihilations via s-channel Higgs boson exchange.
This becomes very efficient when the SM-like Higgs $h$ is on-shell, precluding us from obtaining a relic density matching Planck observations.  
Notice that the second dip is more efficient than the first one, as the Z-mediated dip is momentum suppressed. For heavier $\eta^R$ masses, quartic interactions with gauge bosons become effective.
For $m_{\eta^R} \gsim 80$ GeV, annihilations of $\eta^R$ into $W^+W^-$ and $ZZ$ via quartic couplings are particularly important, thus explaining the third drop in the relic abundance. 
In the mass range $m_{\eta^R}\geq 120$ GeV, $\eta^R$ can annihilate also into two Higgs bosons, $hh$. When $m_{\eta^R}\geq m_t$, a new channel $\eta^R\eta^R\to t\bar{t}$ opens up.  
All these annihilation channels make dark matter annihilation very efficient, and it is difficult to obtain the correct relic density. 
For very heavy $m_{\eta^R}$ the relic density increases due to the suppressed annihilation cross section, which drops as $\sim\frac{1}{m_{\eta^R}^2}$. 
Notice also that the coannihilation channels with $\eta^I$ and $\eta^\pm$ may occur in all regions of the parameter space, with the effect of lowering the relic dark matter density. 

\subsection{Direct detection}

\begin{figure}[h]
\centering
\includegraphics[width=0.3\textwidth]{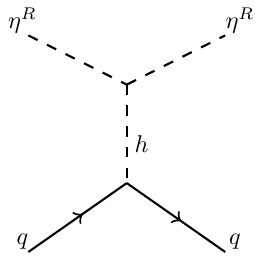}~~
\includegraphics[width=0.3\textwidth]{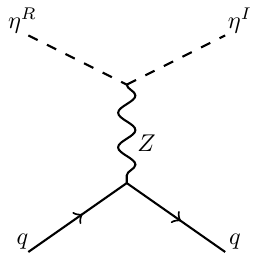}
\caption{Higgs and $Z$-mediated tree-level Feynman diagrams contributing to the elastic scattering of $\eta^R$ off nuclei.}
\label{fig:DD}
\end{figure}

\begin{figure}[h]
\centering
\includegraphics[width=0.8\textwidth]{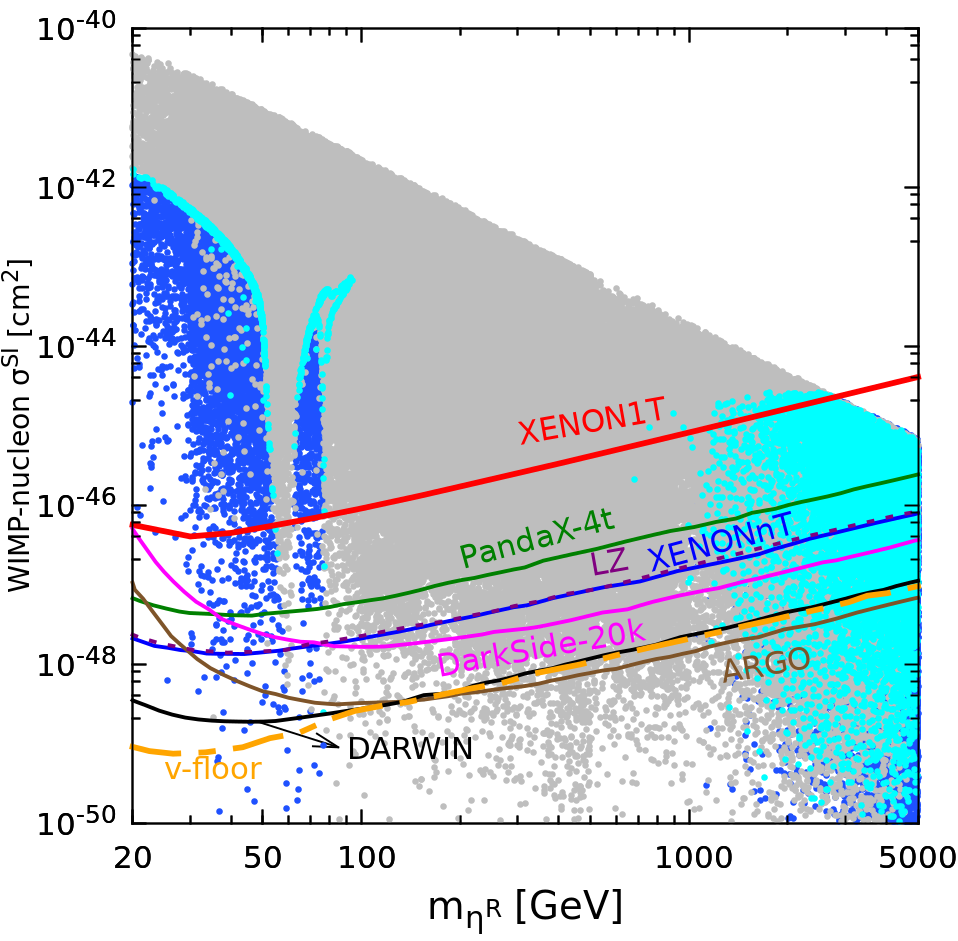}
\caption{\footnotesize{Spin-independent WIMP-nucleon elastic scattering cross section versus the dark matter mass $m_{\eta^R}$, with the same colour code as in Fig.~\ref{fig:Relic}. 
    The solid red line denotes the recent upper bound from the XENON1T experiment~\cite{Aprile:2018dbl}. 
    A broad region of experimentally viable scoto-seesaw model points in cyan fall within future projected sensitivities for the
    PandaX-4t~\cite{Zhang:2018xdp} (green),
    LZ~\cite{Akerib:2018lyp} (dashed purple), XENONnT with 20 ton-yr exposure~\cite{Aprile:2020vtw} (blue), 
    DarkSide-20k~\cite{DS_ESPP} (magenta), DARWIN~\cite{Aalbers:2016jon} (black) and ARGO~\cite{Billard:2021uyg} (brown) proposals.
  The dot-dashed orange line corresponds to the ``neutrino floor'' coming from coherent elastic neutrino scattering~\cite{Billard:2013qya}.
  The upper triangular white region violates perturbativity.
  }}
\label{fig:DD-constraints}
\end{figure}

Let us now study the direct detection prospects of our dark matter $\eta^R$. 
In our model, the tree-level spin-independent $\eta^R$-nucleon cross section is mediated by the Higgs and the $Z$ portals, see Fig.~\ref{fig:DD}. 
%
Notice that, as the $\eta$ doublet has non-zero hypercharge, the $\eta^R$-nucleon spin-independent (SI) cross-section is mediated by the Z-boson.
Generally this exceeds the current limit from direct detection experiments like XENON1T~\cite{Aprile:2018dbl}. 
However this can be easily avoided by taking non-zero $\lambda_5$.
In this case there is a small mass splitting between $\eta^I$ and $\eta^R$, so that the interaction through the Z-boson is kinematically forbiden or leads to inelastic scattering. 
As a result, for nonzero $\lambda_5$, the $\eta^R$-nucleon interaction via the Higgs will be the dominant one.
The coupling between $\eta^R$ and the Higgs boson depends on $\lambda_{345}=\lambda_3+\lambda_4+\lambda_5$ and the $\eta^R$-nucleon cross section is given by
\begin{align}
\sigma^{\text{SI}}=\frac{\lambda_{345}^2}{4\pi m_{h}^4}\frac{m_N^4 f_N^2}{(m_{\eta^R}+m_N)^2}
\end{align} 
where $m_h$ is the mass of SM Higgs boson and $m_N$ is the nucleon mass, i.e. the average of the proton and neutron masses. 
Here $f_N$ is the form factor, which depends on hadronic matrix elements.  
In Fig.~\ref{fig:DD-constraints} we show the spin-independent $\eta^R$-nucleon cross section as a function of the $\eta^R$ mass, for the range of parameters covered by our scan given in Table~\ref{tab:scan}.
%
The color code in Fig.~\ref{fig:DD-constraints} is the same as in Fig.~\ref{fig:Relic}. The red line denotes the latest upper bound from the XENON1T collaboration. 
There are constraints from other experiments as well, such as LUX~\cite{Akerib:2016vxi} and PandaX-II~\cite{Tan:2016zwf}, but weaker when compared to the XENON1T limit. 
We also show the projected sensitivities for the PandaX-4t~\cite{Zhang:2018xdp}, LUX-ZEPLIN(LZ)~\cite{Akerib:2018lyp}, XENONnT~\cite{Aprile:2020vtw}, DarkSide-20k~\cite{DS_ESPP},
DARWIN~\cite{Aalbers:2016jon} and ARGO~\cite{Billard:2021uyg} experiments.
The lower limit corresponding to the ``neutrino floor'' from coherent elastic neutrino scattering is also indicated. 
We see from Fig.~\ref{fig:DD-constraints} that there are low-mass solutions with the correct dark matter relic density.
However, most of these are ruled out by the XENON1T direct detection cross section upper limits.
  Moreover, there are also tight constraints on low mass dark matter from collider searches, as we will discuss in the next section.
\section{Collider constraints}
\label{sec:collider}
In this section we confront our scalar dark matter candidate $\eta^R$ with the latest data from particle colliders, in particular the LHC.
First we note that, if $\eta^R/\eta^I$ are light enough, there are two additional decay channels for the SM-like Higgs boson,
\begin{align}
\Gamma(h\to\eta^R\eta^R)&=\frac{v^2\lambda_{345}^2}{32\pi m_h}\sqrt{1-\frac{4m_{\eta^R}^2}{m_h^2}}\\
\Gamma(h\to\eta^I\eta^I)&=\frac{(m_{\eta^I}^2-m_{\eta^R}^2+\frac{\lambda_{345}}{2}v^2)^2}{8\pi v^2 m_h}\sqrt{1-\frac{4m_{\eta^I}^2}{m_h^2}}
\end{align}
Note that due to LEP limit $m_{\eta^\pm}>m_W$~\cite{Pierce:2007ut}, there is no phase space for the two body decay $h\to\eta^\pm\eta^\pm$. 
The decay mode $\Gamma(h\to\eta^R\eta^R)$ contributes to the invisible Higgs decay width, constrained by the LHC experiments, e.g. the CMS experiment~\cite{Sirunyan:2018owy},
$$\text{BR}(h\to\text{Inv})\leq 0.19.$$ 
The SM-like Higgs boson $h$ also couples to the charged Higgs $\eta^\pm$, contributing to the diphoton decay channel $h\to\gamma\gamma$~\footnote{Note that this invisible Higgs decay and charged scalar contributions to $h\to\gamma\gamma$ are generic features of inert doublet schemes~\cite{Barbieri:2006dq,Cao:2007rm}
as well as scotogenic models~\cite{Bonilla:2016diq,Avila:2019hhv}.}. 
To quantify the deviation from the Standard Model prediction, we define the following parameter 
\begin{align}
R_{\gamma\gamma}=\frac{\text{BR}(h\to\gamma\gamma)}{\text{BR}(h\to\gamma\gamma)^{\text{SM}}}.
\end{align}
The value we use for the Standard Model is $\text{BR}(h\to\gamma\gamma)^{\text{SM}}\approx 2.27\times 10^{-3}$.
\begin{figure}[h]
\centering
\includegraphics[width=0.45\textwidth]{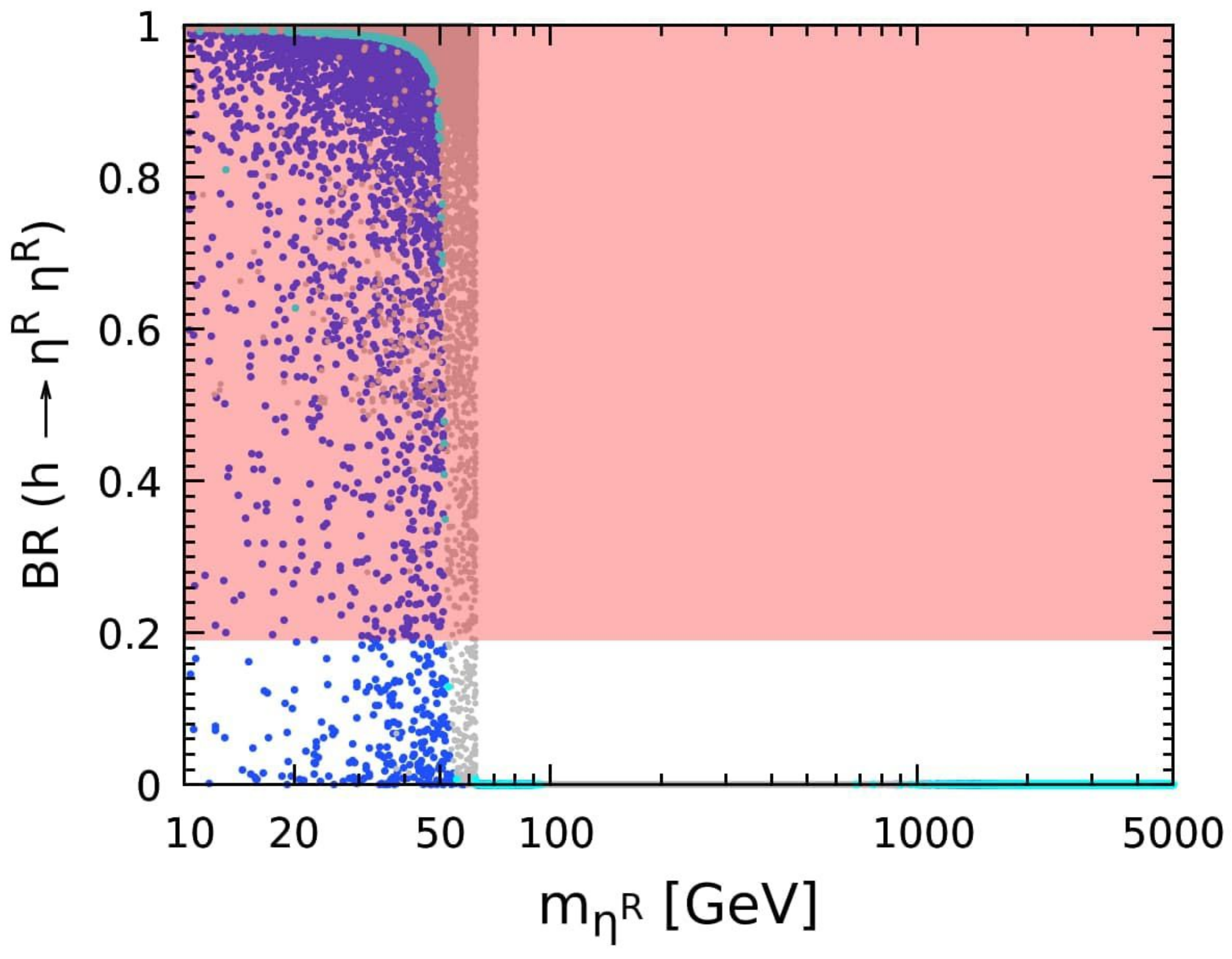}
\includegraphics[width=0.45\textwidth]{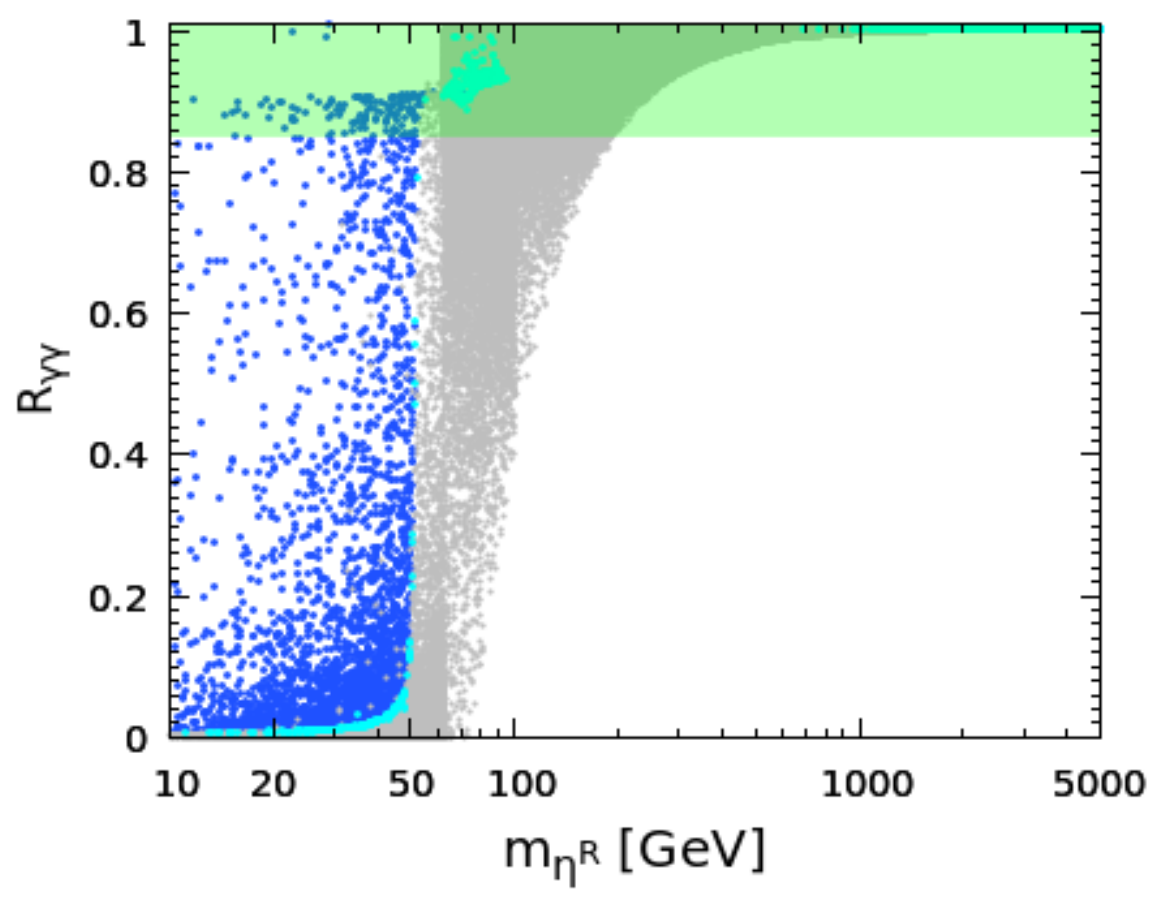}
\caption{Invisible Higgs branching ratio~(left panel) and $R_{\gamma\gamma}$~(right panel) as a function of the dark matter mass $m_{\eta^R}$. The color code is same as in Fig.~\ref{fig:Relic}.
    The shaded red region in the left panel is excluded from the LHC constraint on the invisible Higgs decay~\cite{Sirunyan:2018owy}, while the shaded green band in the right is allowed by
    ATLAS measurements of $R_{\gamma\gamma}$~\cite{Aaboud:2018xdt}.}
\label{fig:higgs-decay}
\end{figure}

The ATLAS and CMS collaborations have studied this decay mode and their combined analysis with 8 TeV data gives $R_{\gamma\gamma}^{\text{exp}}=1.16_{-0.18}^{+0.20}$~\cite{Khachatryan:2016vau}. For the 13 TeV Run-2, there is no combined final data so far, and the available data is separated by production processes~\cite{Aad:2019mbh}. In our analysis we have used 13 TeV ATLAS result which gives the global signal strength measurement of $R_{\gamma\gamma}^{\text{exp}}=0.99_{-0.14}^{+0.15}$~\cite{Aaboud:2018xdt}.
In the left panel of Fig.~\ref{fig:higgs-decay} we show the invisible Higgs branching ratio $\text{BR}(h\to\eta^R\eta^R)$ as a function of the dark matter candidate mass $m_{\eta^R}$.
In the right panel we give the expected $R_{\gamma\gamma}$ values for the same random scan of parameters.
One sees that for low dark matter masses $m_{\eta^R}< 60$~GeV the invisible decay mode $h\to\eta^R\eta^R$ is open and violates the LHC limit $\text{BR}(h\to\text{Inv})\leq 0.19$~\cite{Sirunyan:2018owy}.

Likewise, the $R_{\gamma\gamma}$ measurement rules out the lower dark matter mass region. 
For intermediate dark matter masses in the range $70\,\text{GeV}\leq m_{\eta^R}\leq 100\,\text{GeV}$, there are acceptable solutions with acceptable $R_{\gamma\gamma}\approx 1$. 
In the large mass region $m_{\eta^R}>550$~GeV, the charged Higgs $\eta^{\pm}$ contribution to the diphoton decay mode $h\to\gamma\gamma$ is negligible, so that $R_{\gamma\gamma}$ is close to unity.
From the above discussion, one can say that low mass dark matter with $m_{\eta^R}<60$~GeV is ruled out by LHC constraints.
  However, they can not completely rule out the intermediate mass region $70\,\text{GeV}\leq\,m_{\eta^R}\leq \,100\,\text{GeV}$.
  Moreover, the heavy dark matter mass region $m_{\eta^R}>550$ GeV is completely allowed by LHC constraints.

Before concluding this section we should note that there are also constraints from LEP-I and LEP-II experiments.
  The precise LEP-I measurements rule out SM-gauge bosons decays to dark sector particles~\cite{Cao:2007rm,Gustafsson:2007pc}. This requires that 
\begin{align}
m_{\eta^R}+m_{\eta^I},2m_{\eta^{\pm}}>m_Z,\,\,\text{and}\,\,m_{\eta^R/\eta^I}+m_{\eta^{\pm}}> m_{W}
\end{align}
Although there is no dedicated analysis of LEP-II data in the context of scotogenic dark matter models, Ref.~\cite{Lundstrom:2008ai}
has discussed LEP II limits for the case of the Inert Doublet Model, leading to the limits $m_{\eta^R}\gsim 80$ GeV and $m_{\eta^I}\gsim 100$ GeV and a small $\eta_R-\eta_I$ mass splitting.
Altogether, in view of the above, one can say that intermediate dark matter masses in the range $80\,\text{GeV}\leq m_{\eta^R}\leq 100\,\text{GeV}$ are not inconsistent with collider constraints.
Dark matter heavier than 550 GeV is perfectly allowed. Dark matter masses in between 100 and 550 GeV could also be possible in the presence of another dark matter component.

\black
\section{Electroweak Vacuum Stability}
\label{sec:vacuum stability}
The detailed analysis of the Higgs vacuum within the Standard Model has been carried out in~\cite{Isidori:2001bm,EliasMiro:2011aa,Bezrukov:2012sa,Degrassi:2012ry,Masina:2012tz,Buttazzo:2013uya}. 
Taking into account the updated input top and Higgs boson mass values one finds that the \sm Higgs quartic coupling $\lambda_{\text{SM}}$ becomes negative at $\mu \simeq 10^{10}$ GeV. 
This would imply that the Higgs potential is unbounded from below and the Higgs vacuum is unstable.
A dedicated analysis shows that, actually, the \sm Higgs vacuum is metastable with very long lifetime~\cite{Buttazzo:2013uya}.

Here our aim is to determine the parameter region consistent both with dark matter observations as well as vacuum stability.
We first fix the parameters (quartic couplings and mass of the dark matter candidate) which are consistent with the present day relic density and the direct dark matter detection constraints.  

\begin{figure}[h]
\centering
\includegraphics[width=0.59\textwidth]{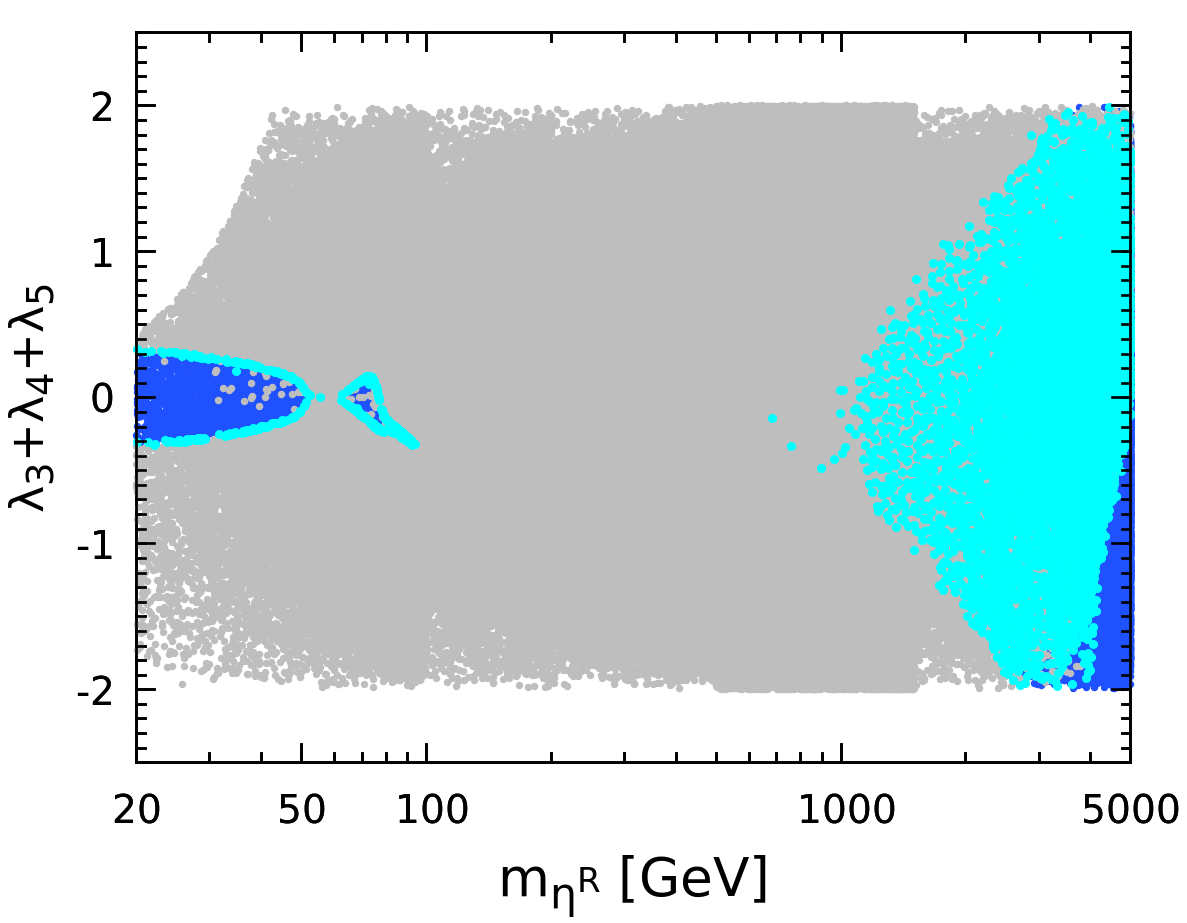}
\caption{Coupling $\lambda_{345}=\lambda_3+\lambda_4+\lambda_5$ as a function of the dark mass $m_{\eta^R}$. The color code is same as in Fig.~\ref{fig:Relic}.}
\label{lambda345}
\end{figure}

In Fig.~\ref{lambda345} we have shown the quartic coupling $\lambda_{345}=\lambda_3+\lambda_4+\lambda_5$ as a function of dark matter mass $m_{\eta^R}$.
%
%
The color code is same as in Fig.~\ref{fig:Relic}. The lower mass region is already excluded by experiment. 
One sees that in the large mass regime above $m_{\eta^R}>550$ GeV, the allowed $\lambda_{345}$ values that successfully explain the relic density while obeying the direct detection limit cover a wide range. 
For relatively large couplings, the evolution of quartic couplings can make them exceed the perturbativity limit even before the Planck scale.
We therefore choose relatively small $\lambda_{3,4,5}$ values. 
This way $\lambda_{345}$ is small enough to match the required relic density and to satisfy the direct detection cross section bound for $m_{\eta^R}>550$ GeV. \\

We now examine the effect of the new particles present in the scoto-seesaw model upon the stability of the electroweak vacuum.
As a ``missing partner'' (3,1) type-I seesaw cloned with the simplest scotogenic sector, the scoto-seesaw contains the a ``right-handed'' neutrino $N$, together with the dark particles $f$ and $\eta$. 
Using SARAH~\cite{Staub:2015kfa} we have computed the two-loop RGEs of the full theory for all the quartic scalar couplings, as well as Yukawa couplings, as given in Appendix~\ref{AppendixA}.

We now summarise our results.
To begin with, in the effective theory where the heavy singlet fermions $N$ and $f$ are integrated out, we have two natural threshold scales $\Lambda_N\approx M_N$ and $\Lambda_f\approx M_f$.
These masses are obtained from Eq.~\eqref{ATM Lagrangian} and \eqref{SOL Lagrangian}. 
Hence, in the RGEs of the full theory, we can simply take the $Y_{N}$ and $Y_{f}$ Yukawa parameters to be given as $\theta\left(\mu-M_{N}\right)Y_{N}$, $\theta\left(\mu-M_{f}\right)Y_{f}$. 
Clearly they do not run in the effective theory.

Our aim is to study the effect of large as well as small Yukawa couplings on vacuum stability. 
As discussed before, we take an adequately small but nonzero value for $\lambda_5$, as required for having a reasonable direct detection cross section. 
For example, with Yukawa couplings $Y_N\sim Y_f\sim\mathcal{O}(1)$ and very large $M_N$, $M_f$ values,
one sees from Eq.~\eqref{atm-sol-mass} that one can easily reproduce the solar and atmospheric scale with $m_\eta^R\sim\mathcal{O}(1\,\text{TeV})$. \\[-.3cm]

\begin{figure}[h]
\centering
\includegraphics[width=0.49\textwidth]{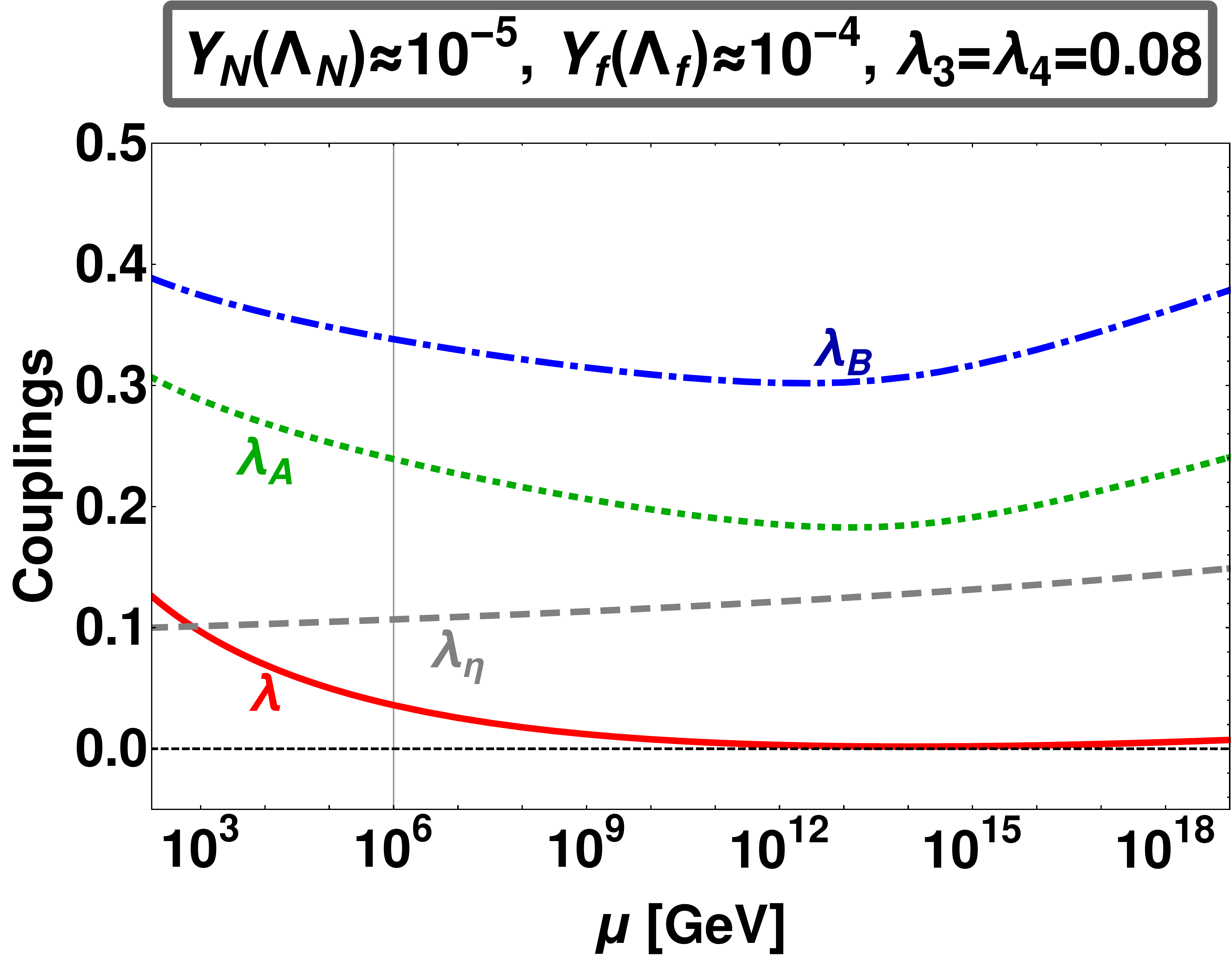}
\includegraphics[width=0.49\textwidth]{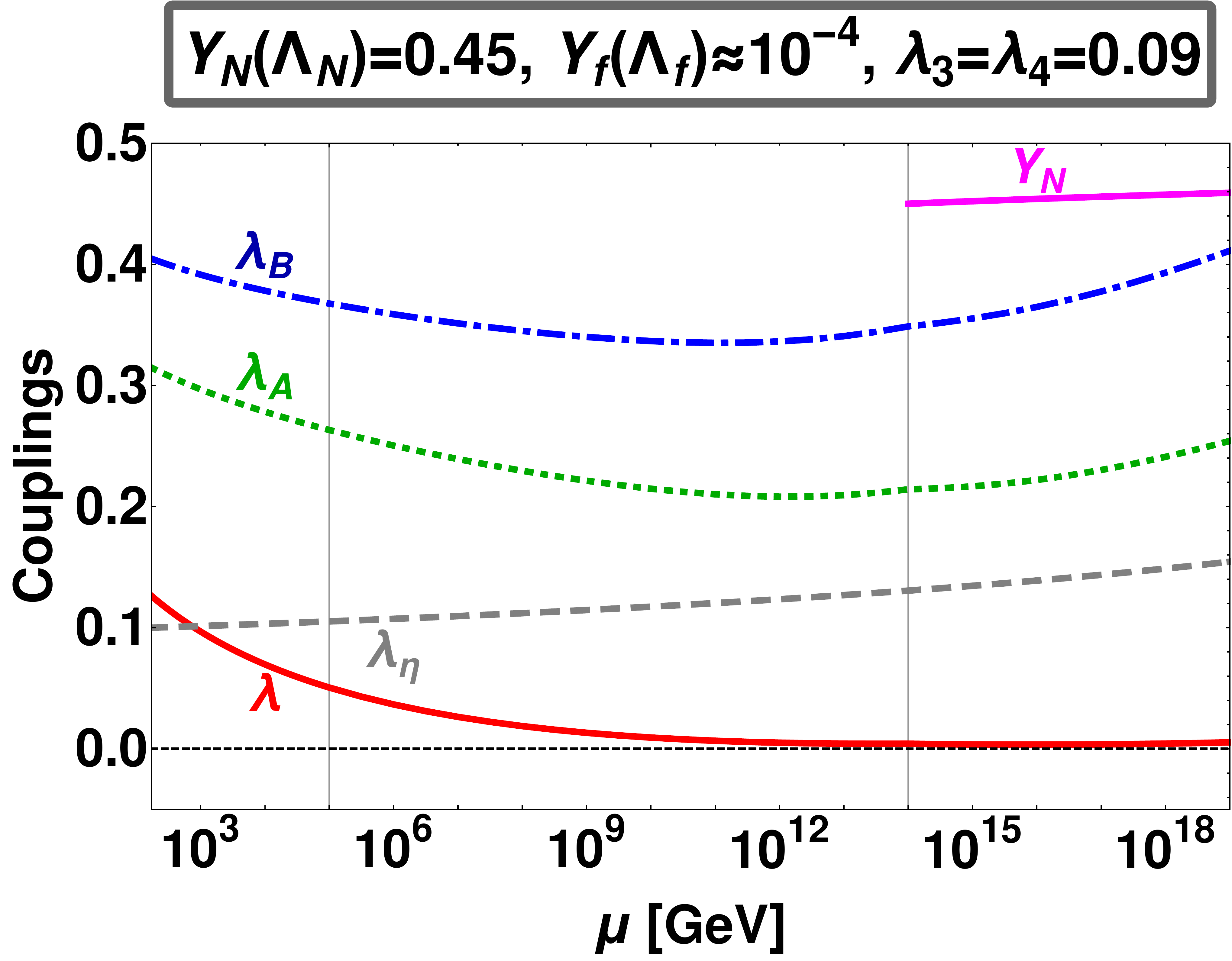}
\includegraphics[width=0.49\textwidth]{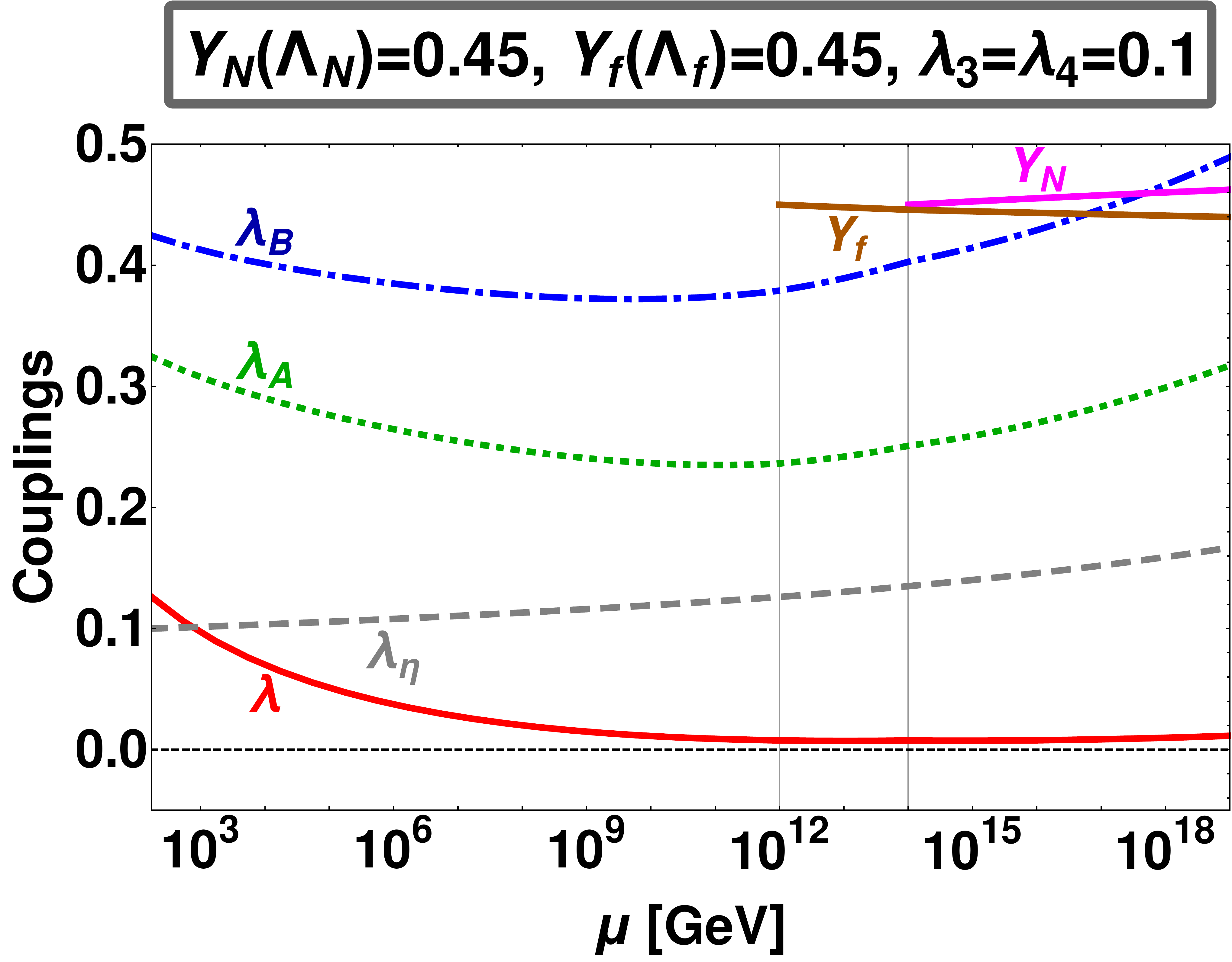}
\caption{\footnotesize{The RGE-evolution of quartic couplings $\lambda$, $\lambda_{\eta}$, $\lambda_A$, $\lambda_B$ defined in Eqs.~(\ref{stability1}),(\ref{stability2}),(\ref{stability3}).
    The Yukawas $Y_N$, $Y_{f}$ are also shown.
    The values given in the boxes are the initial values at the respective threshold scales. All couplings remain perturbative up to the Planck Scale.
    Note that we have fixed $m_{\eta^R} = 1$ TeV and taken small $\lambda_5=10^{-3}$. See text for more details.} }
\label{RGE results}
\end{figure}

We now illustrate in more detail the relevant parameter space of the scoto-seesaw model which is consistent with vacuum stability as well as neutrino and dark matter phenomenology. 
In order to do so we have chosen three sets of benchmarks, given as:
\begin{itemize}
\item  \textbf{BP1:} $M_N\sim 10^{14}$ GeV, $M_f\sim 10^{12}$ GeV, $m_{\eta^R}\sim 10^3$ GeV, $\mathbb{Y}_N\sim 0.45$, $\mathbb{Y}_f\sim 0.45$,
\item \textbf{BP2:}  $M_N\sim 10^{14}$ GeV, $M_f\sim 10^{5}$ GeV, $m_{\eta^R}\sim 10^3$ GeV, $\mathbb{Y}_N\sim 0.45$, $\mathbb{Y}_f\sim 10^{-4}$,
\item \textbf{BP3:} $M_N\sim 10^{6}$ GeV, $M_f\sim 10^{6}$ GeV, $m_{\eta^R}\sim 10^3$ GeV, $\mathbb{Y}_N\sim 10^{-5}$, $\mathbb{Y}_f\sim 10^{-4}$.
\end{itemize}
In the upper left, upper right and bottom panel of Fig.~\ref{RGE results}, we have shown the results for benchmark points \textbf{BP1}, \textbf{BP2} and \textbf{BP3}, respectively. 
We have taken $\lambda_{\eta}=0.1$ and $\lambda=\frac{m_h^2}{2v^2}$ at the electroweak scale. 
Recall that in order to have an absolutely stable vacuum, one needs to satisfy the conditions given in Eqs.~(\ref{stability1}), (\ref{stability2}) and (\ref{stability3}) at all energy scales.\\[-.3cm]

Assuming small $\lambda_5$ we show in Fig.~\ref{RGE results} the evolution of the remaining four quartic couplings $\lambda$, $\lambda_\eta$, $\lambda_A$ and $\lambda_B$. 
%
%
One sees that, with reasonable initial choices, all of the quartic couplings can remain positive and perturbative all the way up to the Planck scale. 
Since the Yukawa couplings have a negative effect in the RGE evolution of $\lambda$, the required value of the quartic couplings $\lambda_3$ and $\lambda_4$ is correspondingly larger,
as seen when going from the upper left to the right panel and finally to the bottom panel in Fig.~\ref{RGE results}. \\[-.3cm] 


To sum up, one sees that the minimal scoto-seesaw model has improved stability properties compared to the type-I seesaw scenario, due to the new scalars needed to realize the scotogenic ``completion''.
The model can explain both solar and atmospheric neutrino mass scales as well as dark matter,
upgrading the (3, 1) type-I seesaw mechanism, which can only generate the atmospheric neutrino mass scale.
Moreover, the minimal scoto-seesaw model leads to a stable and perturbative electroweak vacuum all the way up to the Planck scale.
It can therefore be considered as a full consistent theory for neutrino masses and dark matter.

\section{High energy behavior of the dark parity}
\label{sec:dark-parity}

The conservation of the dark parity is a key feature of the scoto-seesaw model, ensuring dark matter stability as well as the radiative origin of the solar mass scale. 
Without this $\mathbb{Z}_2$ symmetry conservation, the LDP would no longer be stable, and also the solar neutrino splitting would not be ``calculable'' from Eq.~\eqref{atm-sol-mass}. 
It was first pointed out in Ref.~\cite{Merle:2015gea} that renormalization group evolution can alter the scalar potential at high energies, leading to $\mathbb{Z}_2$ breaking. 
It is easy to understand the source of $\mathbb{Z}_2$ symmetry breaking from the one-loop $\beta$ function of the $m_\eta^2$ parameter. 
This is given in~Appendix~\ref{AppendixA}~\cite{Merle:2015ica,Lindner:2016kqk}. 
One needs to focus on the terms which contribute negatively to the evolution of $m_\eta^2$, given in Eq.~\eqref{meta oneloop-running}. 
We see that with relatively large Yukawa coupling $Y_f$ (i.e. $\lambda_5\ll 1$) and $M_f^2> m_\eta^2$, the term $-|M_f|^2|Y_f|^2$ dominates the running of $m_\eta^2$. 
This can quickly drive $m_\eta^2$ towards negative values and induce a minimum of the scalar potential with $\braket{\eta}\neq 0$. 
Notice, however, that there are terms in Eq.~\eqref{meta oneloop-running} which can counter this negative effect.
For example, terms proportional to the quartic scalar couplings $\lambda_3$ and $\lambda_4$ may do so if their signs are properly chosen.
The contribution to the $m_\eta^2$ evolution will be positive for $\lambda_\eta>0$ and $\lambda_{3,4}<0$. 
\begin{figure}[h]
\centering
\includegraphics[width=0.49\textwidth]{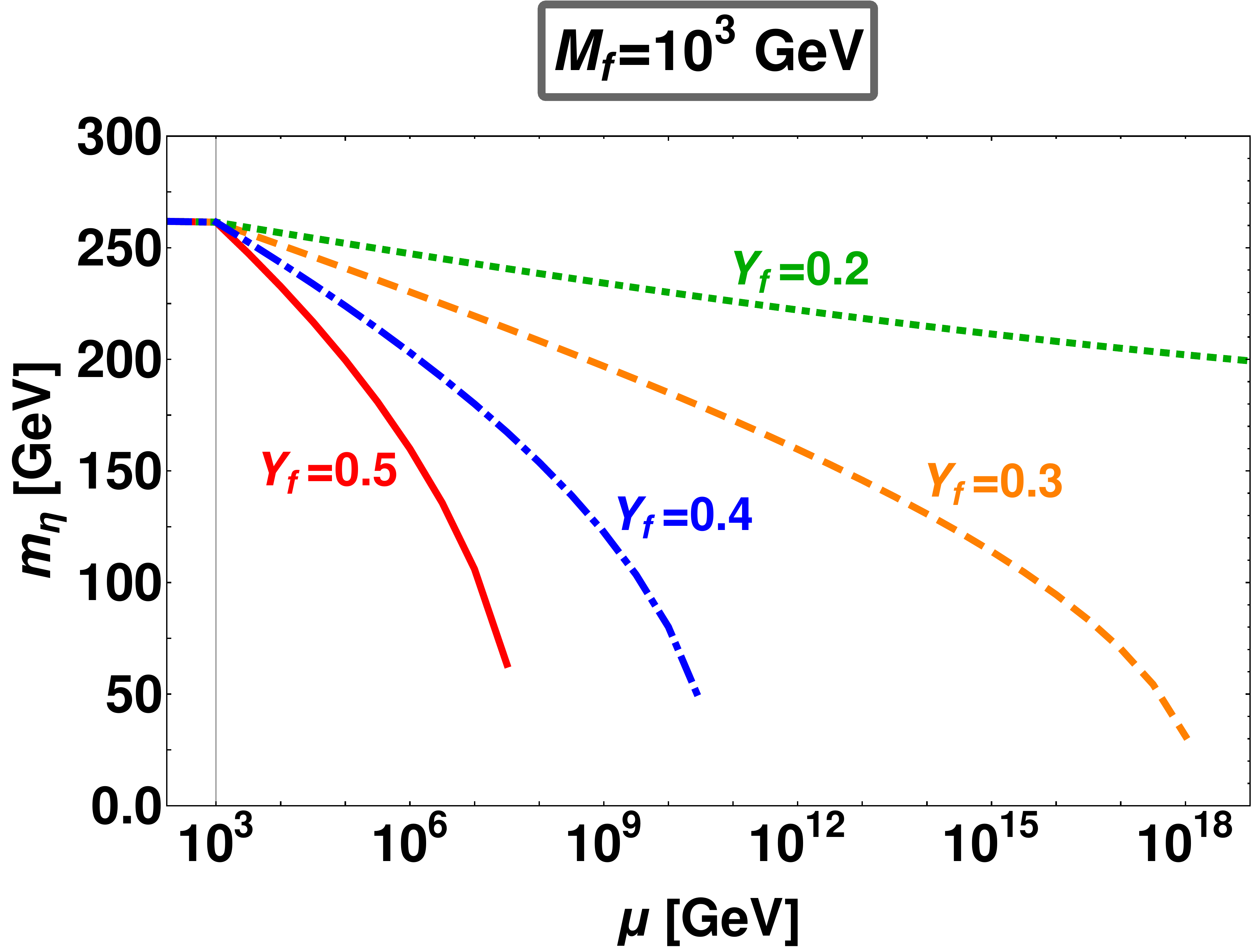}
\includegraphics[width=0.49\textwidth]{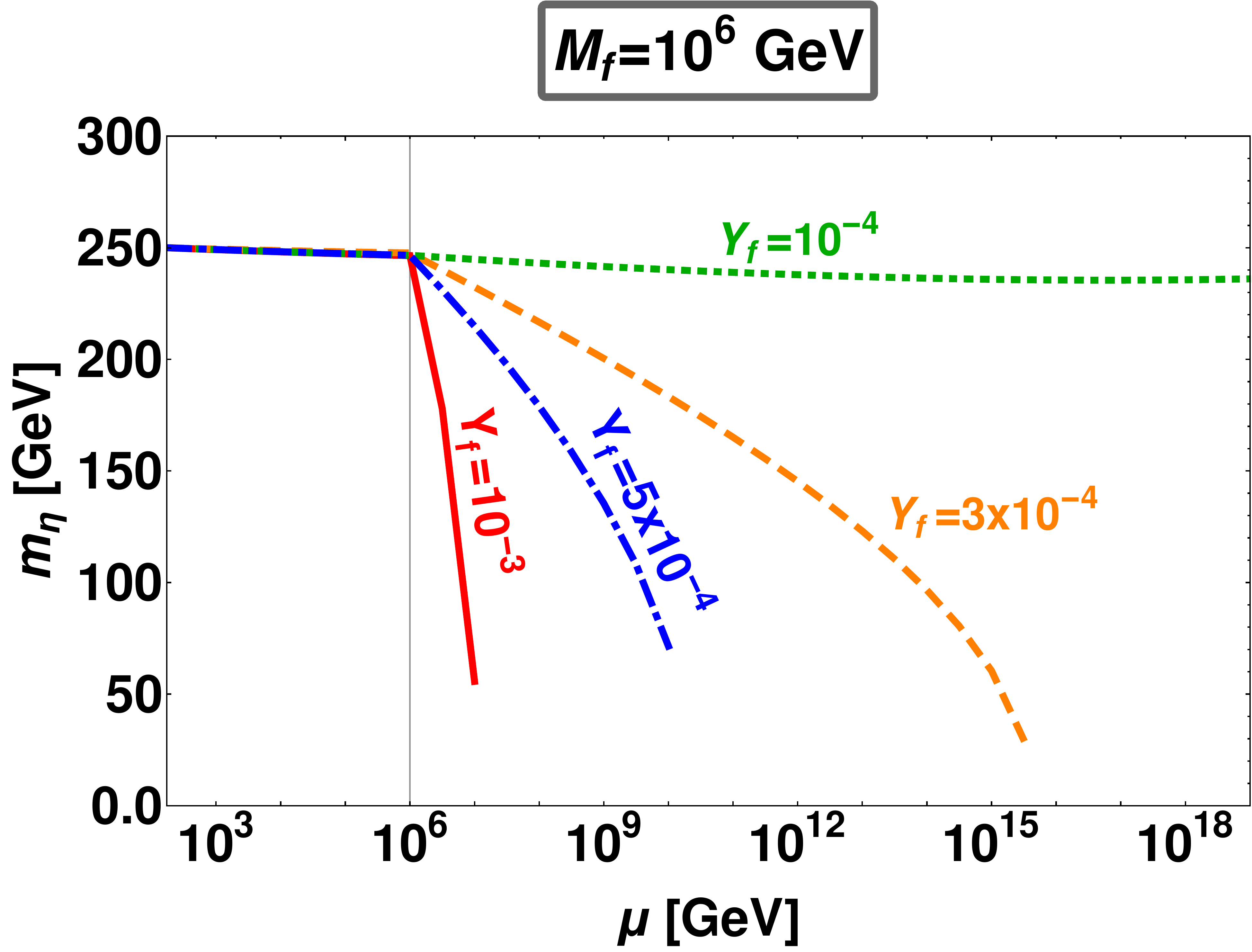}
\caption{\footnotesize{RGE evolution of $m_{\eta}$ as a function of the energy scale $\mu$.
    For both panels we start running at $m_{\eta}=250\,\text{GeV}$, taking $Y_N=0$, $\lambda_3=\lambda_4=0.1$ and $\lambda_5=10^{-3}$.
    Different lines correspond to various Yukawa coupling values, as indicated.
    Note that for $M_f=10^6$GeV (right panel) the Yukawas are much smaller.
    Notice also that all curves terminate when $m_{\eta}^2$ becomes negative ($m_{\eta}$ imaginary) indicating $\mathbb{Z}_2$ breakdown.}}
\label{Z2 breaking}
\end{figure}

Fig.~\ref{Z2 breaking} shows the evolution of the scalar mass $m_{\eta}$ as a function of energy scale $\mu$. 
The results have been obtained for two values of $M_f$, $M_f=10^3$ GeV~(left panel) and $10^6$ GeV~(right panel). 
We have fixed $m_{\eta}^2=250^2\,\text{GeV}^2$, $\lambda_3=\lambda_4=0.1$, $\lambda_5\approx 0$ and $Y_N=0$, for simplicity. 
The red-solid, blue-dot-dashed, orange-dashed and green-dotted lines in the left (right) panel correspond to four values of the Yukawa coupling $Y_f$, as indicated. 
%
As expected, the $\mathbb{Z}_2$ breaking scale decreases for larger $M_f$ due to the effect of the term $-|M_f|^2|Y_f|^2$. 
In other words, the larger the scale $M_f$, the smaller the allowed value of the Yukawa coupling $Y_f$ in order to have the $\mathbb{Z}_2$ symmetry preserved all the way up to the Planck scale. 
From Fig.~\ref{Z2 breaking}, one sees that the allowed value of this Yukawa coupling is $Y_f\leq 0.2$ for $M_f=10^3$ GeV, whereas for $M_f=10^6$ GeV it is $Y_f\leq 10^{-4}$.
Although the different quartic couplings such as $\lambda_3$ and $\lambda_4$ may alter the details, this generic behavior remains.
To sum up, we found that in the scoto-seesaw model the dark parity can be preserved up to the Planck scale over large portions of the parameter space.

\section{Summary and discussion} 
\label{sec:summary-discussion}

We have examined the minimal combination of the the seesaw and scotogenic paradigms for neutrino mass generation and dark matter able to ``explain'' the solar and atmospheric oscillation wavelengths,
  in Eqs.~1 and 2.
The model provides a simple picture where the ``atmospheric'' mass scale arises from the tree-level ``missing partner'' seesaw,
while the ``solar'' scale is induced radiatively by the dark sector, see Fig.~\ref{fig:solar-loop}.
We have derived the full two-loop RGEs for the relevant parameters, such as the quartic Higgs self-coupling $\lambda$ of the Standard Model.
The new scalars present in the scoto-seesaw mechanism improve the stability properties of the electroweak vacuum, as seen in Fig.~\ref{RGE results}. 
We have also explored the consistency of the underlying dark symmetry, as seen in Fig.~\ref{Z2 breaking}.\\[-.4cm]

Concerning phenomenology we have discussed scalar dark matter including the experimental restrictions that follow from colliders, Fig.~\ref{fig:higgs-decay}.
By taking into account the relevant annihiliation channels in Fig.~\ref{fig:annihiliation} we identified viable parameter regions consistent with the required dark matter relic abundance,
Fig.~\ref{fig:Relic}.
Direct dark matter detection by nucleon recoil proceeds through the diagrams in Fig.\ref{fig:DD}.
The expected rates are given in~Fig.~\ref{fig:DD-constraints} and offer promising results for upcoming dark matter experiments.
We found that the low dark matter mass region $m_{\eta^R}\lsim 60$~GeV is ruled out by LHC data, but the intermediate region $80\,\text{GeV}\,\lsim m_{\eta^R} \lsim 100\,\text{GeV}$ is still allowed
  both by LHC and LEP data. The heavier mass region $m_{\eta^R} \gsim 550$ GeV is free from collider constraints. \\[-.4cm]

Our construction is very attractive from the point of view of neutrino physics.
  In contrast to the original scotogenic model where 2 (or 3) species of either dark fermion or dark scalars are needed to generate masses for 2 (or 3) neutrinos,
  our dark sector is truly minimal, with only one dark fermion and one dark scalar.
  Therefore, the allowed parameter space differs from the original scotogenic model, though these differences do not translate into a phenomenological smoking-gun signature
 which can easily distinguish it from canonical scotogenic model in Ref.~\cite{Ma:2006km}.\\[-.4cm]
  
Before closing we also note an important phenomenological implication of the minimal scoto-seesaw model that can make it testable.
Namely, it implies that one of the neutrinos is massless, as can be readily seen from Eq.~\eqref{eq:mnu-scoto-seesaw}.
This leads to a lower bound on the \znbb decay rates even for a normal-ordered neutrino mass spectrum~\cite{Reig:2018ztc,Barreiros:2018bju,Leite:2019grf,Avila:2019hhv}.
As shown in Fig.2 of Ref.~\cite{Avila:2019hhv}, for an inverted mass spectrum the \0nbb lower bound lies substantially higher than in the generic case for that ordering.
As a result it falls within the sensitivity of future experiments such as nEXO~\cite{Jewell:2020ceq}.
In the scoto-seesaw one has that a positive \0nbb decay discovery would open bright prospects for the underpinning of the value of the relevant elusive Majorana phase. 
As discussed in~\cite{Rojas:2018wym}, the expected rates for \lfv processes can also lie within reach of experiments, providing additional signatures. 
In summary, the scoto-seesaw model is a theoretically consistent and phenomenologicaly interesting ``dark matter completion'' of the type-I seesaw mechanism.

\begin{acknowledgments}
  This work is supported by the Spanish grant FPA2017-85216-P (AEI/FEDER, UE), PROMETEO/2018/165 (Generalitat Valenciana).
  R.S. is supported by SERB, Government of India grant SRG/2020/002303.
\end{acknowledgments}

\appendix
\section{Scalar dark matter annihilation mechanisms}
\label{AppendixB}

In the minimal scoto-seesaw model the relic abundance of the lightest dark particle $\eta^R$ is determined by the following annihilation and coannihilation diagrams.
\begin{figure}[h]
\centering
\includegraphics[width=0.31\textwidth,height=6cm,keepaspectratio]{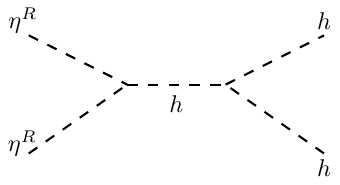}
\includegraphics[width=0.31\textwidth,height=6cm,keepaspectratio]{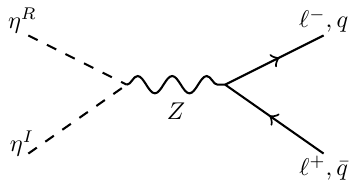}
\includegraphics[width=0.31\textwidth]{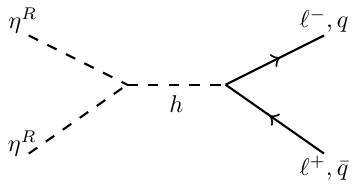}\\
\includegraphics[width=0.31\textwidth]{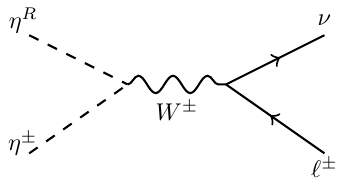}
\includegraphics[width=0.31\textwidth]{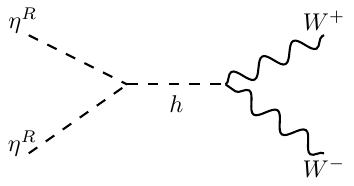}
\includegraphics[width=0.31\textwidth]{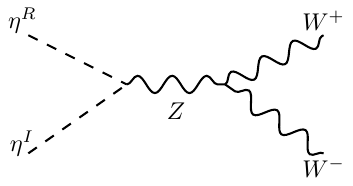} \\
\includegraphics[width=0.24\textwidth,height=4cm,keepaspectratio]{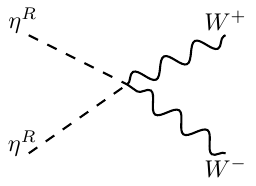}
\includegraphics[width=0.24\textwidth,height=4cm,keepaspectratio]{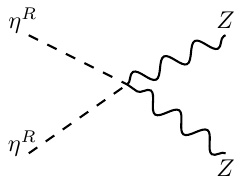}
\includegraphics[width=0.24\textwidth,height=4cm,keepaspectratio]{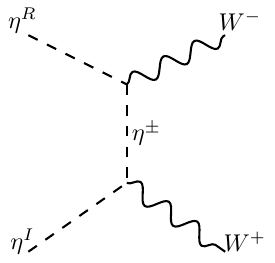} 
\includegraphics[width=0.24\textwidth,height=4cm,keepaspectratio]{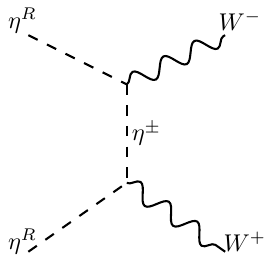}
\includegraphics[width=0.25\textwidth,height=4cm,keepaspectratio]{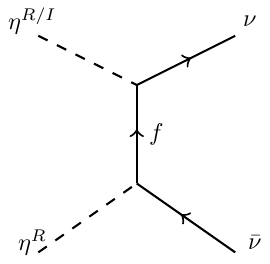}
\includegraphics[width=0.25\textwidth,height=4cm,keepaspectratio]{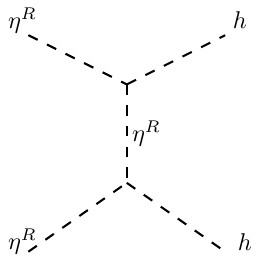} 
\includegraphics[width=0.25\textwidth,height=4cm,keepaspectratio]{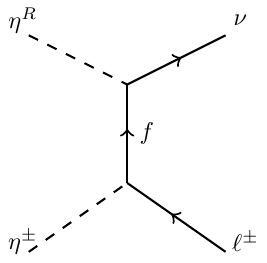}
\caption{\footnotesize{Annihilation and coannihilation diagrams contributing to the relic abundance of $\eta^R$.}}
\label{fig:annihiliation}
\end{figure}

\section{Renormalization group equations}
\label{AppendixA}

The evolution of a given parameter c in the theory is described by the appropriate $\beta$ function, given by,
\begin{align*}
 \frac{dc}{dt}\equiv\beta_{c}=\frac{1}{16\pi^{2}}\beta_{c}^{(1)}+\frac{1}{(16\pi^{2})^{2}}\beta_{c}^{(2)} \, .
\end{align*}
where $\beta_{c}^{(1)}$ are the one-loop renormalization group (RG) coefficients, while $\beta_{c}^{(2)}$ correspond to the two-loop RG corrections. 
\subsection{Higgs quartic scalar self coupling}
\label{app:scoto-quartic}
The scalar potential of the scoto-seesaw model is given in Eq.~\eqref{eq:scoto-pot}.
The model contains five quartic couplings $\lambda, \lambda_\eta, \lambda_3, \lambda_4, \lambda_5$.
The one-loop and two-loop RG equations of the Higgs quartic self-coupling $\lambda$ are given by 
\begin{align}
\beta_{\lambda}^{(1)} & =  
+\frac{27}{200} g_{1}^{4} +\frac{9}{20} g_{1}^{2} g_{2}^{2} +\frac{9}{8} g_{2}^{4} +2 \lambda_{3}^{2} +2 \lambda_3 \lambda_4 +\lambda_{4}^{2} + \lambda_{5}^{2} 
-\frac{9}{5} g_{1}^{2} \lambda -9 g_{2}^{2} \lambda +24 \lambda^{2} \nonumber \\ 
 &+12 \lambda y_{t}^{2} +4 \lambda \mbox{Tr}\Big({Y_N  Y_N^{\dagger}}\Big) 
 -6 y_{t}^{4} -2 \mbox{Tr}\Big({Y_N  Y_N^{\dagger}  Y_N  Y_N^{\dagger}}\Big) \, ,
 \end{align}
 \begin{align}
 \beta_{\lambda}^{(2)} & =  
-\frac{3537}{2000} g_{1}^{6} -\frac{1719}{400} g_{1}^{4} g_{2}^{2} -\frac{303}{80} g_{1}^{2} g_{2}^{4} +\frac{291}{16} g_{2}^{6}
+\frac{9}{10} g_{1}^{4} \lambda_3 +\frac{15}{2} g_{2}^{4} \lambda_3 +\frac{12}{5} g_{1}^{2} \lambda_{3}^{2} +12 g_{2}^{2} \lambda_{3}^{2} \nonumber \\
&-8 \lambda_{3}^{3} 
 +\frac{9}{20} g_{1}^{4} \lambda_4 +\frac{3}{2} g_{1}^{2} g_{2}^{2} \lambda_4 +\frac{15}{4} g_{2}^{4} \lambda_4 +\frac{12}{5} g_{1}^{2} \lambda_3 \lambda_4 +12 g_{2}^{2} \lambda_3 \lambda_4 -12 \lambda_{3}^{2} \lambda_4 +\frac{6}{5} g_{1}^{2} \lambda_{4}^{2} \nonumber \\ 
 &+3 g_{2}^{2} \lambda_{4}^{2} -16 \lambda_3 \lambda_{4}^{2} -6 \lambda_{4}^{3} -\frac{3}{5} g_{1}^{2} \lambda_{5}^{2} -20 \lambda_3 \lambda_{5}^{2} -22 \lambda_4 \lambda_{5}^{2} +\frac{1953}{200} g_{1}^{4} \lambda +\frac{117}{20} g_{1}^{2} g_{2}^{2} \lambda \nonumber \\
  &-\frac{51}{8} g_{2}^{4} \lambda -20 \lambda_{3}^{2} \lambda -20 \lambda_3 \lambda_4 \lambda -12 \lambda_{4}^{2} \lambda -14 \lambda_{5}^{2} \lambda 
 +\frac{108}{5} g_{1}^{2} \lambda^{2} +108 g_{2}^{2} \lambda^{2} -312 \lambda^{3} \nonumber \\
  & -4 \lambda_{3}^{2} \mbox{Tr}\Big({Y_f  Y_{f}^{\dagger}}\Big)  
 -4 \lambda_3 \lambda_4 \mbox{Tr}\Big({Y_f  Y_{f}^{\dagger}}\Big) -2 \lambda_{4}^{2} \mbox{Tr}\Big({Y_f  Y_{f}^{\dagger}}\Big)
 -2 \lambda_{5}^{2} \mbox{Tr}\Big({Y_f  Y_{f}^{\dagger}}\Big) -\frac{171}{100} g_{1}^{4} y_{t}^{2} \nonumber \\
 &+\frac{63}{10} g_{1}^{2} g_{2}^{2} y_{t}^{2} -\frac{9}{4} g_{2}^{4} y_{t}^{2} +\frac{17}{2} g_{1}^{2} \lambda y_{t}^{2} +\frac{45}{2} g_{2}^{2} \lambda y_{t}^{2} 
 +80 g_{3}^{2} \lambda y_{t}^{2} -144 \lambda^{2} y_{t}^{2} -\frac{9}{100} g_{1}^{4} \mbox{Tr}\Big({Y_N  Y_N^{\dagger}}\Big) \nonumber \\
 &-\frac{3}{10} g_{1}^{2} g_{2}^{2} \mbox{Tr}\Big({Y_N  Y_N^{\dagger}}\Big) 
 -\frac{3}{4} g_{2}^{4} \mbox{Tr}\Big({Y_N  Y_N^{\dagger}}\Big) +\frac{3}{2} g_{1}^{2} \lambda \mbox{Tr}\Big({Y_N  Y_N^{\dagger}}\Big) 
 +\frac{15}{2} g_{2}^{2} \lambda \mbox{Tr}\Big({Y_N  Y_N^{\dagger}}\Big) \nonumber \\
& -48 \lambda^{2} \mbox{Tr}\Big({Y_N  Y_N^{\dagger}}\Big) 
 -3 \lambda \mbox{Tr}\Big({Y_f  Y_{f}^{\dagger}  Y_N  Y_N^{\dagger}}\Big) -\frac{8}{5} g_{1}^{2} y_{t}^{4} 
 -32 g_{3}^{2} y_{t}^{4} -3 \lambda y_{t}^{4}
 - \lambda \mbox{Tr}\Big({Y_N  Y_N^{\dagger}  Y_N  Y_N^{\dagger}}\Big) \nonumber \\
& +30 y_{t}^{6}  +2 \mbox{Tr}\Big({Y_f  Y_{f}^{\dagger}  Y_N  Y_N^{\dagger}  Y_N  Y_N^{\dagger}}\Big)  
 +10 \mbox{Tr}\Big({Y_N  Y_N^{\dagger}  Y_N  Y_N^{\dagger}  Y_N  Y_N^{\dagger}}\Big) \, .
 \end{align}
\begin{align}
\beta_{\lambda_\eta}^{(1)} & =  
+\frac{27}{200} g_{1}^{4} +\frac{9}{8} g_{2}^{4} +2 \lambda_{3}^{2} +2 \lambda_3 \lambda_4 +\lambda_{4}^{2}+\lambda_{5}^{2}+\frac{9}{20} g_{1}^{2} \Big(-4 \lambda_\eta  + g_{2}^{2}\Big)-9 g_{2}^{2} \lambda_\eta +24 \lambda_{\eta}^{2} \nonumber \\
& +4 \lambda_\eta \text{Tr}({Y_f  Y_f^{\dagger}}) -2 \text{Tr}(\Big({Y_f  Y_f^{\dagger}Y_f Y_f^{\dagger}}\Big)\Big) 
\end{align}

\begin{align}
\beta_{\lambda_\eta}^{(2)} & =  
-\frac{3537}{2000} g_{1}^{6} -\frac{1719}{400} g_{1}^{4} g_{2}^{2} -\frac{303}{80} g_{1}^{2} g_{2}^{4} +\frac{291}{16} g_{2}^{6} +\frac{9}{10} g_{1}^{4} \lambda_3 +\frac{15}{2} g_{2}^{4} \lambda_3 +\frac{12}{5} g_{1}^{2} \lambda_{3}^{2} +12 g_{2}^{2} \lambda_{3}^{2} \nonumber \\
& -8 \lambda_{3}^{3}  +\frac{9}{20} g_{1}^{4} \lambda_4 +\frac{3}{2} g_{1}^{2} g_{2}^{2} \lambda_4 +\frac{15}{4} g_{2}^{4} \lambda_4 +\frac{12}{5} g_{1}^{2} \lambda_3 \lambda_4 +12 g_{2}^{2} \lambda_3 \lambda_4 -12 \lambda_{3}^{2} \lambda_4 +\frac{6}{5} g_{1}^{2} \lambda_{4}^{2} \nonumber \\ 
 &+3 g_{2}^{2} \lambda_{4}^{2} -16 \lambda_3 \lambda_{4}^{2} -6 \lambda_{4}^{3} -\frac{3}{5} g_{1}^{2} \lambda_{5}^{2} -20 \lambda_3 \lambda_{5}^{2} -22 \lambda_4 \lambda_{5}^{2} +\frac{1953}{200} g_{1}^{4} \lambda_\eta +\frac{117}{20} g_{1}^{2} g_{2}^{2} \lambda_\eta \nonumber \\ 
 &-\frac{51}{8} g_{2}^{4} \lambda_\eta -20 \lambda_{3}^{2} \lambda_\eta -20 \lambda_3 \lambda_4 \lambda_\eta -12 \lambda_{4}^{2} \lambda_\eta -14 \lambda_{5}^{2} \lambda_\eta +\frac{108}{5} g_{1}^{2} \lambda_{\eta}^{2} +108 g_{2}^{2} \lambda_{\eta}^{2} -312 \lambda_{\eta}^{3} \nonumber \\ 
 &- \lambda_\eta \text{Tr}\Big({Y_f  Y_f^{\dagger}Y_f Y_f^{\dagger}}\Big) +10 \text{Tr}(\Big({Y_f  Y_f^{\dagger}}Y_f Y_f^{\dagger}Y_f Y_f^{\dagger}\Big) -3 \lambda_\eta \text{Tr}\Big({Y_f  Y_N^{\dagger}} {Y_N  Y_f^{\dagger}}\Big) -4 \lambda_{3}^{2} \text{Tr}\Big({Y_N  Y_N^{\dagger}}\Big) \nonumber \\
& -4 \lambda_3 \lambda_4 \text{Tr}\Big({Y_N  Y_N^{\dagger}}\Big) 
-2 \lambda_{4}^{2} \text{Tr}\Big({Y_N  Y_N^{\dagger}}\Big) -2 \lambda_{5}^{2} \text{Tr}\Big({Y_N  Y_N^{\dagger}}\Big)  +\text{Tr}\Big({Y_f  Y_f^{\dagger}}\Big) \Big( 2 \text{Tr}\Big({Y_f  Y_N^{\dagger}} {Y_N  Y_f^{\dagger}}\Big) \nonumber \\
&  -\frac{3}{100} \Big(10 g_{1}^{2} \Big(-5 \lambda_\eta  + g_{2}^{2}\Big) 
+ 25 \Big(-10 g_{2}^{2} \lambda_\eta  + 64 \lambda_{\eta}^{2}  + g_{2}^{4}\Big) + 3 g_{1}^{4} \Big)\Big) -12 \lambda_{3}^{2} y_t^2 \nonumber \\
&-12 \lambda_3 \lambda_4 y_t^2 -6 \lambda_{4}^{2} y_t^2 -6 \lambda_{5}^{2} y_t^2 
\end{align} 
\begin{align}
\beta_{\lambda_4}^{(1)} & =  
+\frac{9}{5} g_{1}^{2} g_{2}^{2} -\frac{9}{5} g_{1}^{2} \lambda_4 -9 g_{2}^{2} \lambda_4 +8 \lambda_3 \lambda_4 +4 \lambda_{4}^{2} +8 \lambda_{5}^{2} +4 \lambda_4 \lambda_\eta +4 \lambda_4 \lambda +2 \lambda_4 \text{Tr}\Big({Y_f  Y_f^{\dagger}}\Big) \nonumber \\ 
 &-4 \text{Tr}\Big({Y_f  Y_N^{\dagger}}{Y_N  Y_f^{\dagger}}\Big) +2 \lambda_4 \text{Tr}\Big({Y_N  Y_N^{\dagger}}\Big) +6 \lambda_4 y_t^2 
\end{align}
\begin{align}
\beta_{\lambda_4}^{(2)} & =  
-\frac{657}{50} g_{1}^{4} g_{2}^{2} -\frac{42}{5} g_{1}^{2} g_{2}^{4} +\frac{6}{5} g_{1}^{2} g_{2}^{2} \lambda_3 +\frac{1413}{200} g_{1}^{4} \lambda_4 +\frac{153}{20} g_{1}^{2} g_{2}^{2} \lambda_4 -\frac{231}{8} g_{2}^{4} \lambda_4 +\frac{12}{5} g_{1}^{2} \lambda_3 \lambda_4 \nonumber \\ 
 &+36 g_{2}^{2} \lambda_3 \lambda_4 -28 \lambda_{3}^{2} \lambda_4 +\frac{24}{5} g_{1}^{2} \lambda_{4}^{2} +18 g_{2}^{2} \lambda_{4}^{2} -28 \lambda_3 \lambda_{4}^{2} +\frac{48}{5} g_{1}^{2} \lambda_{5}^{2} +54 g_{2}^{2} \lambda_{5}^{2} -48 \lambda_3 \lambda_{5}^{2} \nonumber \\ 
 &-26 \lambda_4 \lambda_{5}^{2} +6 g_{1}^{2} g_{2}^{2} \lambda_\eta +\frac{24}{5} g_{1}^{2} \lambda_4 \lambda_\eta -80 \lambda_3 \lambda_4 \lambda_\eta -40 \lambda_{4}^{2} \lambda_\eta -48 \lambda_{5}^{2} \lambda_\eta -28 \lambda_4 \lambda_{\eta}^{2} +6 g_{1}^{2} g_{2}^{2} \lambda \nonumber \\
& +\frac{24}{5} g_{1}^{2} \lambda_4 \lambda -80 \lambda_3 \lambda_4 \lambda -40 \lambda_{4}^{2} \lambda -48 \lambda_{5}^{2} \lambda -28 \lambda_4 \lambda^{2} -\frac{9}{2} \lambda_4 \text{Tr}\Big({Y_f  Y_f^{\dagger}Y_f Y_f^{\dagger}}\Big) +\text{Tr}\Big(Y_N Y_N^{\dagger}\Big) \nonumber \\
&\Big(-\frac{3}{5} g_{1}^{2} g_{2}^{2} +\frac{3}{4} g_{1}^{2} \lambda_4 +\frac{15}{4} g_{2}^{2} \lambda_4  -8 \lambda_3 \lambda_4 -4 \lambda_{4}^{2} 
 -8 \lambda_{5}^{2}  -8 \lambda_4 \lambda \Big) -\frac{9}{2} \lambda_4 \text{Tr}\Big({Y_N  Y_N^{\dagger} Y_N Y_N^{\dagger}}\Big) \nonumber \\
& +\text{Tr}\Big({Y_f  Y_N^{\dagger}}{Y_N  Y_f^{\dagger}}\Big) \Big(-3 \lambda_4  + 8 \lambda_3  + 8 \text{Tr}\Big({Y_N  Y_N^{\dagger}}\Big) \Big) + \text{Tr}\Big({Y_f  Y_f^{\dagger}}\Big) \Big(-\frac{3}{5} g_{1}^{2} g_{2}^{2} +\frac{3}{4} g_{1}^{2} \lambda_4 \nonumber \\
&+\frac{15}{4} g_{2}^{2} \lambda_4  -8 \lambda_3 \lambda_4 -4 \lambda_{4}^{2} -8 \lambda_{5}^{2} -8 \lambda_4 \lambda_\eta +8 \text{Tr}\Big({Y_f  Y_N^{\dagger}}{Y_N  Y_f^{\dagger}}\Big) \Big)
  +y_t^2\Big( \frac{63}{5} g_{1}^{2} g_{2}^{2}  +\frac{17}{4} g_{1}^{2} \lambda_4  \nonumber \\ 
 &+\frac{45}{4} g_{2}^{2} \lambda_4  +40 g_{3}^{2} \lambda_4 -24 \lambda_3 \lambda_4  -12 \lambda_{4}^{2}  -24 \lambda_{5}^{2}  -24 \lambda_4 \lambda -\frac{27}{2} \lambda_4 y_t^2  \Big)  
\end{align}
\begin{align}
\beta_{\lambda_3}^{(1)} & =  
+\frac{27}{100} g_{1}^{4} -\frac{9}{10} g_{1}^{2} g_{2}^{2} +\frac{9}{4} g_{2}^{4} -\frac{9}{5} g_{1}^{2} \lambda_3 -9 g_{2}^{2} \lambda_3 +4 \lambda_{3}^{2} +2 \lambda_{4}^{2} +2 \lambda_{5}^{2} +12 \lambda_3 \lambda_\eta \nonumber \\
&+4 \lambda_4 \lambda_\eta +12 \lambda_3 \lambda +4 \lambda_4 \lambda +2 \lambda_3 \text{Tr}\Big({Y_f  Y_f^{\dagger}}\Big) +2 \lambda_3 \text{Tr}\Big({Y_N  Y_N^{\dagger}}\Big) +6 \lambda_3 y_t^2
\end{align}

\begin{align}
\beta_{\lambda_3}^{(2)} & =  
-\frac{3537}{1000} g_{1}^{6} +\frac{909}{200} g_{1}^{4} g_{2}^{2} +\frac{33}{40} g_{1}^{2} g_{2}^{4} +\frac{291}{8} g_{2}^{6} +\frac{1773}{200} g_{1}^{4} \lambda_3 +\frac{33}{20} g_{1}^{2} g_{2}^{2} \lambda_3 -\frac{111}{8} g_{2}^{4} \lambda_3 +\frac{6}{5} g_{1}^{2} \lambda_{3}^{2} \nonumber \\ 
 &+6 g_{2}^{2} \lambda_{3}^{2} -12 \lambda_{3}^{3} +\frac{9}{10} g_{1}^{4} \lambda_4 -\frac{9}{5} g_{1}^{2} g_{2}^{2} \lambda_4 +\frac{15}{2} g_{2}^{4} \lambda_4 -12 g_{2}^{2} \lambda_3 \lambda_4 -4 \lambda_{3}^{2} \lambda_4 -\frac{6}{5} g_{1}^{2} \lambda_{4}^{2} \nonumber \\ 
 &+6 g_{2}^{2} \lambda_{4}^{2} -16 \lambda_3 \lambda_{4}^{2} -12 \lambda_{4}^{3} +\frac{12}{5} g_{1}^{2} \lambda_{5}^{2} -18 \lambda_3 \lambda_{5}^{2} -44 \lambda_4 \lambda_{5}^{2} +\frac{27}{10} g_{1}^{4} \lambda_\eta -3 g_{1}^{2} g_{2}^{2} \lambda_\eta \nonumber \\ 
 &+\frac{45}{2} g_{2}^{4} \lambda_\eta +\frac{72}{5} g_{1}^{2} \lambda_3 \lambda_\eta +72 g_{2}^{2} \lambda_3 \lambda_\eta -72 \lambda_{3}^{2} \lambda_\eta +\frac{24}{5} g_{1}^{2} \lambda_4 \lambda_\eta +36 g_{2}^{2} \lambda_4 \lambda_\eta -32 \lambda_3 \lambda_4 \lambda_\eta \nonumber \\ 
 &-28 \lambda_{4}^{2} \lambda_\eta -36 \lambda_{5}^{2} \lambda_\eta -60 \lambda_3 \lambda_{\eta}^{2} -16 \lambda_4 \lambda_{\eta}^{2} +\frac{27}{10} g_{1}^{4} \lambda -3 g_{1}^{2} g_{2}^{2} \lambda +\frac{45}{2} g_{2}^{4} \lambda +\frac{72}{5} g_{1}^{2} \lambda_3 \lambda \nonumber \\ 
 &+72 g_{2}^{2} \lambda_3 \lambda -72 \lambda_{3}^{2} \lambda +\frac{24}{5} g_{1}^{2} \lambda_4 \lambda +36 g_{2}^{2} \lambda_4 \lambda -32 \lambda_3 \lambda_4 \lambda -28 \lambda_{4}^{2} \lambda -36 \lambda_{5}^{2} \lambda -60 \lambda_3 \lambda^{2} \nonumber \\ 
 &-16 \lambda_4 \lambda^{2} -\frac{9}{2} \lambda_3 \text{Tr}\Big({Y_f  Y_f^{\dagger}} Y_f Y_f^{\dagger}\Big) -\frac{9}{100} g_{1}^{4} \text{Tr}\Big({Y_N  Y_N^{\dagger}}\Big) +\frac{3}{10} g_{1}^{2} g_{2}^{2} \text{Tr}\Big({Y_N  Y_N^{\dagger}}\Big) -\frac{3}{4} g_{2}^{4} \text{Tr}\Big({Y_N  Y_N^{\dagger}}\Big) \nonumber \\ 
 &+\frac{3}{4} g_{1}^{2} \lambda_3 \text{Tr}\Big({Y_N  Y_N^{\dagger}}\Big) +\frac{15}{4} g_{2}^{2} \lambda_3 \text{Tr}\Big({Y_N  Y_N^{\dagger}}\Big) -4 \lambda_{3}^{2} \text{Tr}\Big({Y_N  Y_N^{\dagger}}\Big) -2 \lambda_{4}^{2} \text{Tr}\Big({Y_N  Y_N^{\dagger}}\Big) -2 \lambda_{5}^{2} \text{Tr}\Big({Y_N  Y_N^{\dagger}}\Big) \nonumber \\ 
 &-24 \lambda_3 \lambda \text{Tr}\Big({Y_N  Y_N^{\dagger}}\Big) -8 \lambda_4 \lambda \text{Tr} \Big({Y_N  Y_N^{\dagger}}\Big) -\frac{9}{2} \lambda_3 \text{Tr}\Big({Y_N  Y_N^{\dagger}}Y_N Y_N^{\dagger}\Big) + \text{Tr}\Big({Y_f  Y_f^{\dagger}}\Big) \Big(-\frac{9}{100} g_{1}^{4} \nonumber \\
&+\frac{3}{10} g_{1}^{2} g_{2}^{2} -\frac{3}{4} g_{2}^{4} +\frac{3}{4} g_{1}^{2} \lambda_3 +\frac{15}{4} g_{2}^{2} \lambda_3 -4 \lambda_{3}^{2} -2 \lambda_{4}^{2} -2 \lambda_{5}^{2} -24 \lambda_3 \lambda_\eta -8 \lambda_4 \lambda_\eta +4 \text{Tr}\Big({Y_f  Y_N^{\dagger}}{Y_N  Y_f^{\dagger}}\Big)  \Big)\nonumber \\ 
 &+\text{Tr}\Big({Y_f  Y_N^{\dagger}}\Big) \Big( \text{Tr}\Big({Y_N  Y_f^{\dagger}}\Big) \Big(-3 \lambda_3  + 4 \text{Tr}\Big({Y_N  Y_N^{\dagger}}\Big)  + 8 \lambda_4 \Big)\Big) 
 +y_t^2\Big(-\frac{171}{100} g_{1}^{4}  -\frac{63}{10} g_{1}^{2} g_{2}^{2}  -\frac{9}{4} g_{2}^{4}\nonumber \\
&  +\frac{17}{4} g_{1}^{2} \lambda_3 
 +\frac{45}{4} g_{2}^{2} \lambda_3  +40 g_{3}^{2} \lambda_3  -12 \lambda_{3}^{2}  -6 \lambda_{4}^{2}  -6 \lambda_{5}^{2}  -72 \lambda_3 \lambda  -24 \lambda_4 \lambda    -\frac{27}{2} \lambda_3 y_t^2 \Big)
\end{align}
\begin{align}
\beta_{\lambda_5}^{(1)} & =  
-\frac{9}{5} g_{1}^{2} \lambda_5 -9 g_{2}^{2} \lambda_5 +8 \lambda_3 \lambda_5 +12 \lambda_4 \lambda_5 +4 \lambda_5 \lambda_\eta +4 \lambda_5 \lambda +2 \lambda_5 \text{Tr}\Big({Y_f  Y_f^{\dagger}}\Big) +2 \lambda_5 \text{Tr}\Big({Y_N  Y_N^{\dagger}}\Big)\nonumber \\
& +6 \lambda_5 y_t^2
\end{align}

\begin{align}
\beta_{\lambda_5}^{(2)} & =  
+\frac{1413}{200} g_{1}^{4} \lambda_5 +\frac{57}{20} g_{1}^{2} g_{2}^{2} \lambda_5 -\frac{231}{8} g_{2}^{4} \lambda_5 +\frac{48}{5} g_{1}^{2} \lambda_3 \lambda_5 +36 g_{2}^{2} \lambda_3 \lambda_5 -28 \lambda_{3}^{2} \lambda_5 +\frac{72}{5} g_{1}^{2} \lambda_4 \lambda_5 \nonumber \\ 
 &+72 g_{2}^{2} \lambda_4 \lambda_5 -76 \lambda_3 \lambda_4 \lambda_5 -32 \lambda_{4}^{2} \lambda_5 +6 \lambda_{5}^{3} -\frac{12}{5} g_{1}^{2} \lambda_5 \lambda_\eta -80 \lambda_3 \lambda_5 \lambda_\eta -88 \lambda_4 \lambda_5 \lambda_\eta -28 \lambda_5 \lambda_{\eta}^{2} \nonumber \\ 
 &-\frac{12}{5} g_{1}^{2} \lambda_5 \lambda -80 \lambda_3 \lambda_5 \lambda -88 \lambda_4 \lambda_5 \lambda -28 \lambda_5 \lambda^{2} +\frac{1}{4} \lambda_5 \Big(15 g_{2}^{2}  -16 \Big(2 \lambda_3  + 2 \lambda_\eta  + 3 \lambda_4 \Big) + 3 g_{1}^{2} \Big)\text{Tr}\Big({Y_f  Y_f^{\dagger}}\Big) \nonumber \\ 
 &-\frac{1}{2} \lambda_5 \text{Tr}\Big({Y_f  Y_f^{\dagger}Y_f Y_f^{\dagger}}\Big) -3 \lambda_5 \text{Tr}\Big({Y_f  Y_N^{\dagger}}{Y_N  Y_f^{\dagger}}\Big) +\frac{3}{4} g_{1}^{2} \lambda_5 \text{Tr}\Big({Y_N  Y_N^{\dagger}}\Big) +\frac{15}{4} g_{2}^{2} \lambda_5 \text{Tr}\Big({Y_N  Y_N^{\dagger}}\Big) \nonumber \\
& -8 \lambda_3 \lambda_5 \text{Tr}\Big({Y_N  Y_N^{\dagger}}\Big)  
 -12 \lambda_4 \lambda_5 \text{Tr}\Big({Y_N  Y_N^{\dagger}}\Big) -8 \lambda_5 \lambda \text{Tr}\Big({Y_N  Y_N^{\dagger}}\Big) -\frac{1}{2} \lambda_5 \text{Tr}\Big({Y_N  Y_N^{\dagger}}Y_N Y_N^{\dagger}\Big)  \nonumber \\ 
 &+y_t^2\Big(\frac{17}{4} g_{1}^{2} \lambda_5  +\frac{45}{4} g_{2}^{2} \lambda_5  +40 g_{3}^{2} \lambda_5  -24 \lambda_3 \lambda_5  
 -36 \lambda_4 \lambda_5  -24 \lambda_5 \lambda   -\frac{3}{2} \lambda_5 y_t^2\Big)
\end{align}
 \subsection{Yukawa Couplings}
 \label{app:scoto-yuk}
 The one-loop and two-loop RG equations for the Yukawa couplings $Y_f$, $Y_N$ and $y_t$ are given by 
 \begin{align}
 \beta_{Y_f}^{(1)} & =  
\frac{1}{20} \Big(10 \left(3 {Y_f  Y_{f}^{\dagger}  Y_f} + {Y_N  Y_N^{\dagger}  Y_f}\right) + Y_f \Big(20 \mbox{Tr}\Big({Y_f  Y_{f}^{\dagger}}\Big)  -9 \Big(5 g_{2}^{2}  + g_{1}^{2}\Big)\Big)\Big)\, ,
 \end{align}
 \begin{align}
\beta_{Y_f}^{(2)} & =  
+\frac{1}{80} \Big(33 g_{1}^{2} {Y_N  Y_N^{\dagger}  Y_f} +165 g_{2}^{2} {Y_N  Y_N^{\dagger}  Y_f} -160 \lambda_3 {Y_N  Y_N^{\dagger}  Y_f} 
-320 \lambda_4 {Y_N  Y_N^{\dagger}  Y_f} \nonumber \\ 
 & +120 {Y_f  Y_{f}^{\dagger}  Y_f  Y_{f}^{\dagger}  Y_f} -20 {Y_f  Y_{f}^{\dagger}  Y_N  Y_N^{\dagger}  Y_f} 
 -20 {Y_N  Y_N^{\dagger}  Y_N  Y_N^{\dagger}  Y_f} -180 {Y_N  Y_N^{\dagger}  Y_f} y_{t}^{2} \nonumber \\ 
 &+3 {Y_f  Y_{f}^{\dagger}  Y_f} \Big(225 g_{2}^{2}  -320 \lambda_\eta  -60 \mbox{Tr}\Big({Y_f  Y_{f}^{\dagger}}\Big)  + 93 g_{1}^{2} \Big)  
 +Y_f \Big(\frac{117}{200} g_{1}^{4} -\frac{27}{20} g_{1}^{2} g_{2}^{2} \nonumber \\
 &-\frac{21}{4} g_{2}^{4} 
 +\lambda_{3}^{2}+\lambda_3 \lambda_4 +\lambda_{4}^{2}+\frac{3}{2} \lambda_{5}^{2} +6 \lambda_{\eta}^{2} +\frac{3}{8} \Big(5 g_{2}^{2}  + g_{1}^{2}\Big)\mbox{Tr}\Big({Y_f  Y_{f}^{\dagger}}\Big) \nonumber \\ 
 & -\frac{9}{4} \mbox{Tr}\Big({Y_f  Y_{f}^{\dagger}  Y_f  Y_{f}^{\dagger}}\Big) -\frac{3}{4} \mbox{Tr}\Big({Y_f  Y_{f}^{\dagger}  Y_N  Y_N^{\dagger}}\Big) \Big) \, ,
 \end{align}
 \begin{align}
 \beta_{Y_N}^{(1)} & =  
+\frac{1}{2} \Big( 3 {Y_N  Y_N^{\dagger}  Y_N}  + {Y_f  Y_{f}^{\dagger}  Y_N}\Big)
 +Y_N \Big(3 y_{t}^{2}  -\frac{9}{20} g_{1}^{2}  -\frac{9}{4} g_{2}^{2} + \mbox{Tr}\Big({Y_N  Y_N^{\dagger}}\Big)\Big)\, ,
  \end{align}
 \begin{align} 
\beta_{Y_N}^{(2)} & =  
+\frac{1}{80} \Big(279 g_{1}^{2} {Y_N  Y_N^{\dagger}  Y_N} +675 g_{2}^{2} {Y_N  Y_N^{\dagger}  Y_N} -960 \lambda {Y_N  Y_N^{\dagger}  Y_N} 
  -20 {Y_f  Y_{f}^{\dagger}  Y_f  Y_{f}^{\dagger}  Y_N}  \nonumber \\ 
 &-20 {Y_N  Y_N^{\dagger}  Y_f  Y_{f}^{\dagger}  Y_N} +120 {Y_N  Y_N^{\dagger}  Y_N  Y_N^{\dagger}  Y_N}  
 +{Y_f  Y_{f}^{\dagger}  Y_N} \Big(-160 \lambda_3  + 165 g_{2}^{2}  -320 \lambda_4 \nonumber \\
 &+ 33 g_{1}^{2}  -60 \mbox{Tr}\Big({Y_f  Y_{f}^{\dagger}}\Big) \Big)
 -540 {Y_N  Y_N^{\dagger}  Y_N} y_{t}^{2} -180 {Y_N  Y_N^{\dagger}  Y_N} \mbox{Tr}\Big({Y_N  Y_N^{\dagger}}\Big) \nonumber \\ 
 &+Y_N \Big(\frac{117}{200} g_{1}^{4} -\frac{27}{20} g_{1}^{2} g_{2}^{2} -\frac{21}{4} g_{2}^{4} +\lambda_{3}^{2}+\lambda_3 \lambda_4 +\lambda_{4}^{2}+\frac{3}{2} \lambda_{5}^{2} 
 +6 \lambda^{2} 
 +\frac{17}{8} g_{1}^{2} y_{t}^{2} \nonumber \\
 &+\frac{45}{8} g_{2}^{2} y_{t}^{2} +20 g_{3}^{2} y_{t}^{2} 
 +\frac{3}{8} g_{1}^{2} \mbox{Tr}\Big({Y_N  Y_N^{\dagger}}\Big) +\frac{15}{8} g_{2}^{2} \mbox{Tr}\Big({Y_N  Y_N^{\dagger}}\Big) 
 -\frac{3}{4} \mbox{Tr}\Big({Y_f  Y_{f}^{\dagger}  Y_N  Y_N^{\dagger}}\Big) \nonumber \\ 
 &-\frac{27}{4} y_{t}^{4} -\frac{9}{4} \mbox{Tr}\Big({Y_N  Y_N^{\dagger}  Y_N  Y_N^{\dagger}}\Big) \Big) \, ,
 \end{align}
 \begin{align}
 \beta_{y_{t}}^{(1)} & =  
\frac{3}{2} y_{t}^{3}  
 +y_{t} \Big( 3 y_{t}^{2}  -8 g_{3}^{2}  -\frac{17}{20} g_{1}^{2}  -\frac{9}{4} g_{2}^{2} + \mbox{Tr}\Big({Y_N  Y_N^{\dagger}}\Big)\Big) \, ,
 \end{align}
 \begin{align} 
\beta_{y_t}^{(2)} & =  
+\frac{1}{80} \Big(120 y_{t}^{5}
 +y_{t}^{3} \Big(1280 g_{3}^{2}    -180 \mbox{Tr}\Big({Y_N  Y_N^{\dagger}}\Big)  + 223 g_{1}^{2} -540 y_{t}^{2}  + 675 g_{2}^{2}  -960 \lambda \Big)\Big ) \nonumber \\ 
 &+y_{t} \Big(\frac{1267}{600} g_{1}^{4} -\frac{9}{20} g_{1}^{2} g_{2}^{2} -\frac{21}{4} g_{2}^{4} +\frac{19}{15} g_{1}^{2} g_{3}^{2} +9 g_{2}^{2} g_{3}^{2} -108 g_{3}^{4} 
 +\lambda_{3}^{2}+\lambda_3 \lambda_4 +\lambda_{4}^{2}+\frac{3}{2} \lambda_{5}^{2} \nonumber \\
 &+6 \lambda^{2}
  +\frac{17}{8} g_{1}^{2} y_{t}^{2} +\frac{45}{8} g_{2}^{2}y_{t}^{2}
 +20 g_{3}^{2} y_{t}^{2} +\frac{3}{8} g_{1}^{2} \mbox{Tr}\Big({Y_N  Y_N^{\dagger}}\Big) +\frac{15}{8} g_{2}^{2} \mbox{Tr}\Big({Y_N  Y_N^{\dagger}}\Big)\nonumber \\ 
 &-\frac{3}{4} \mbox{Tr}\Big({Y_f  Y_{f}^{\dagger}  Y_N  Y_N^{\dagger}}\Big) -\frac{27}{4} y_{t}^{4} -\frac{9}{4} \mbox{Tr}\Big({Y_N  Y_N^{\dagger}  Y_N  Y_N^{\dagger}}\Big) \Big)
 \end{align}
\subsection{Scalar Mass term}
\label{sec:scalar mass term}
The evolution of the scalar mass-squared term ${m_\eta^2}$ is dictated by the $\beta$ function
\begin{align} 
\beta_{m_\eta^2}^{(1)} & =  
12 \lambda_\eta m_\eta^2  + 2 \left(-2 |M_f|^2  + m_\eta^2\right)\text{Tr}\left({Y_f^{\dagger}  Y_f}\right)  -2 (\lambda_4  + 2 \lambda_3) \mu_H^2   \nonumber \\
& - \left(\frac{9}{10} g_{1}^{2}  + \frac{9}{2} g_{2}^{2}\right) m_\eta^2 
\label{meta oneloop-running}
\end{align}

\bibliographystyle{utphys}
\bibliography{bibliography} 
\end{document}